\documentclass[
 reprint,
 superscriptaddress,
 amsmath,amssymb,
 aps, physrev,
floatfix,
]{revtex4-2}

\usepackage{graphicx}
\usepackage{dcolumn}
\usepackage{bm}
\usepackage{hyperref}

\usepackage{color}

\definecolor{mypurple}{rgb}{0.5, 0, 0.85}
\definecolor{myblue}{rgb}{0, 0.2, 0.85}
\definecolor{mygreen}{rgb}{0.0, 0.42, 0.24}
\definecolor{mypink}{rgb}{0.79, 0, 0.39}

\newcommand{\chimera}{\textsc{Chimera}}
\newcommand{\msun}{\ensuremath{M_\odot}}

\begin{document}

\title{\textbf{Gravitational Waves as a Probe of Core Collapse Supernova Progenitor Structure} 
}

\author{R. Daniel Murphy}
\affiliation{
Department of Physics and Astronomy, University of Tennessee, Knoxville, TN 37996-1200, USA
}
\email{Contact author:rmurph16@vols.utk.edu}

\author{Elle Brinkman}
\affiliation{
Department of Physics and Astronomy, University of Tennessee, Knoxville, TN 37996-1200, USA
}

\author{Colter J. Richardson}
\affiliation{
Department of Physics and Astronomy, University of Tennessee, Knoxville, TN 37996-1200, USA
}

\author{Evan Semenak}
\affiliation{
Department of Physics and Astronomy, University of Tennessee, Knoxville, TN 37996-1200, USA
}

\author{Anthony Mezzacappa}
\affiliation{
Department of Physics and Astronomy, University of Tennessee, Knoxville, TN 37996-1200, USA
}

\author{Pedro Marronetti}
\affiliation{
Physics Division, National Science Foundation, Alexandria, Virginia 22314, USA
}

\author{Eric J. Lentz}
\affiliation{
Department of Physics and Astronomy, University of Tennessee, Knoxville, TN 37996-1200, USA
}
\affiliation{
Physics Division, Oak Ridge National Laboratory, P.O. Box 2008, Oak Ridge, TN 37831-6354, USA
}

\author{Stephen W. Bruenn}
\affiliation{
Department of Physics, Florida Atlantic University, Boca Raton, FL 33431-0991, USA
}

\date{\today}

\begin{abstract}
We present the gravitational wave predictions from two-dimensional core collapse supernova (CCSN) simulations initiated from two nearly identical progenitors that have significantly different internal structures due to their late-stage stellar evolution. At the time of collapse, the 15.78 \msun\ and 15.79 \msun\ progenitors have compactness parameters $\xi_{2.5}$ of 0.136 and 0.206, respectively. We connect several features of the gravitational wave signal from each model to its previously explored explosion dynamics. In particular, the greater accretion onto the PNS of the more compact model is evident in broad-band frequency features with larger amplitude gravitational wave strains and greater gravitational wave energy release when compared to the less compact model. Additionally, the faster contraction rate of the more compact model is reflected in the $\sim$26\% greater slope of the $g$-/$f$-mode feature (gfF) evolution of the gravitational wave signal. This work shows that in principle gravitational wave detection may provide information about interior stellar structure.
\end{abstract}

\maketitle

\section{Introduction}
\label{sec:introduction}

As we anticipate the next Galactic core collapse supernova (CCSN), it is necessary to continue investigations into the effects progenitor properties have on the CCSN-generated gravitational wave signal. Already, studies have shown intrinsic differences in gravitational wave signals from progenitors with differing mass \cite{AnMuMu17,VaBuRa19a,MeMaLa23,VaBuWa23}, rotation rate \cite{AnMuJa19,PaCoPa19,PoMu20,PaLiCo21,TaKoFo21,ZhOc22,WaPa24}, and magnetic field strengths \cite{JaPoMu22,BuGuFo23,PoMuAg23,PoMu24}. Beyond these physical effects on the gravitational wave signal, uncertainties in our understanding of fundamental physics also influences the signals generated in CCSN simulations, as seen in simulations that implement differing nuclear equations of state\cite{MaJaMu09,MoRaBu18,PaLiCo18,PoMuHe21,AnZhSi21}, differing neutrino properties \cite{MoTaKo25,EhAbJa24}, and that include quantum phase transitions \cite{YaKoHa07,ZhOcCh20,KuFiTa22,ZhOc22}. Understanding these effects is not only theoretically important to build our understanding of the gravitational wave production; it allows for the possibility of extracting progenitor parameters \cite{WaCoOc20,AlfeBrown_21,PaWaCo21}, proto-neutron star (PNS) parameters \cite{BiMaTo21,PoMu22,BrBiOb23,WoFrMi23,SaSaDo24}, and constraining fundamental physics theories---e.g., the nuclear equation of state \cite{JaMuHe23,MuCaMe24}---through a gravitational wave detection. In this study, we aim to both improve our understanding of gravitational wave production in CCSNe and to explore the use of a gravitational wave detection as a means to investigate stellar structure, by presenting the gravitational wave signals of two progenitors with nearly identical mass but differing internal structure. The explosion dynamics of these 15.78-\msun\ and 15.79-\msun\ progenitors are described in \citet{BrSiLe23}. 

Throughout the CCSN event, gravitational waves will be produced deep within and around the PNS evolving beneath the shockwave created at core bounce. For rotating progenitors, at core bounce there is a significant time-dependent quadrupole mass moment, which produces a prominent spike in the gravitational wave strain, with secondary spikes then produced from bounce-induced oscillations of the PNS collectively termed the ``bounce signal" \cite{DiOtMa08}. Immediately after bounce, the presence of large gradients in entropy and electron fraction causes prompt convection within the PNS \cite{BrMe94}, which produces broad-band gravitational waves between 5 and 20 ms after bounce \cite{MaJaMu09}. After these large gradients are removed through prompt convection, the gravitational wave signal is relatively quiet as the PNS is further deleptonized by neutrino diffusion, although \citet{MeMaLa23} point out that this quiet period is likely an artifact of initiating a simulation with a spherically symmetric progenitor. Approximately 100 ms after bounce, sustained neutrino emission has created lepton gradients deep within the PNS large enough to cause Ledoux convection, which will produce high-frequency gravitational waves \cite{MuRaBu04}. The lepton gradients are sustained on the order of seconds as neutrinos continue to diffuse out of the PNS, thus sustaining the Ledoux convection as well. Outside the surface of the PNS, but below the shock radius, there exists a net neutrino heating layer where neutrino-driven convection becomes turbulent and produces low-frequency ($<250$ Hz) gravitational waves \cite{MuOtBu09}. In addition to neutrino-driven convection, the Standing Accretion Shock Instability (SASI) produced by non-spherical perturbations at the shock interface, also produces low-frequency gravitational waves \cite{KoOhYa07}. Large-scale accretion funnels can develop at every stage postbounce and impinge upon the PNS surface to create high-frequency gravitational waves as well, though the onset of explosion will slow this accretion and its resulting excitation of gravitational waves. Finally, both the anisotropic emission of neutrinos \cite{BuHa96} and the asymmetrical explosion \cite{RiZaAn22}  will cause low-frequency, time-dependent quadrupole moments that result in very-low-frequency gravitational waves ($<50$ Hz), often referred to as gravitational wave memory, though the memory applies only to the final state. For reviews on the generation of gravitational waves in CCSN see \citet{AbPaRa_2022_book} and \citet{MeZa_24}.

At every phase of the CCSN, gravitational waves thus carry information imprinted by the dynamics of the explosion, which in turn depends, in part, on the internal structure of the progenitor star. In previous studies, varying the compactness parameter of the progenitor accounts for the bulk internal structural differences between progenitor models. The compactness parameter for the inner core with mass $M$ is defined by \citet{OcOt11} at the time of core bounce as 
\begin{equation}
    \xi_M\equiv\frac{M/M_\odot}{R(M)/1000 \ \rm km}, \label{eq:compact}
\end{equation}
and it has been shown to quantitatively affect the gravitational wave signal produced in CCSN simulations. \citet{WaCoOc20} used one-dimensional CCSN simulations to show a positive correlation between the compactness of progenitors that failed to explode and the rate of change---i.e., slope with respect to time---of the peak frequency of the gravitational wave emission. The peak gravitational wave emission frequency is approximated through a combination of PNS modal analysis following \citet{MoRaBu18} and the peak frequency evolution assuming a $g-$mode type oscillation as derived by \citet{MuJaMa13}. Using two-dimensional CCSN simulations, \citet{PaWaCo21} corroborated this positive correlation with direct determinations of the slope of the peak frequency evolution from gravitational wave calculations. For rotating CCSNe, they showed the feasibility of determining the compactness of the progenitor core using this peak frequency slope and the bounce signal. \citet{WaPa24} saw the same positive correlation between compactness parameter and peak frequency evolution for two-dimensional CCSN simulations with progenitors of differing internal structure arising from binary accretion during stellar evolution.

Our study contributes to the growing body of gravitational waveforms generated from CCSN simulations, and investigates the gravitational wave production in detail not only from two progenitors with varying compactness parameters---i.e., varying internal structures---but also from progenitors that are nearly identical in every other respect. This allows us to show, in a limited case, the effects of varying internal structure in isolation of other progenitor parameters. The general differences we observe can be attributed in part to the compactness parameter, as in the studies cited above, but we will also endeavor to connect specific gravitational wave features to the unique dynamics described in the CCSN simulations of \citet{BrSiLe23}.

This paper is organized as follows: Section \ref{sec:models_and_methods} provides a brief overview of our core collapse supernova models and details our gravitational wave extraction method; Section \ref{sec:results} presents the gravitational wave analysis for each model; Section \ref{sec:summary} summarizes our results and discusses their implications.

\section{Models and Methods}
\label{sec:models_and_methods}

\subsection{Core-Collapse Supernova Models}
\label{sec:ccsn_models}
We examine the gravitational wave emission from two-dimensional simulations of collapse, and explosion, of two progenitors of nearly equal mass, performed with the \chimera\ simulation code \cite{BrBlHi20}. The zero age main sequence masses of the progenitors are 15.78 \msun\ and 15.79 \msun, and the models were evolved to precollapse by \citet{SuWoHe18} and updated as described in Section 2 of \citet{BrSiLe23}. The explosion dynamics of each model are described in detail in Section 5 of \citet{BrSiLe23}. The following figures from \citet{BrSiLe23} are of particular importance to our study: Figure 4, showing the compactness of each model, Figure 5, showing the shock radius (top) and mass accretion rate at the shock location (bottom), Figure 8, showing the PNS mass and radius evolution, and Figure 16, showing the neutrino accretion and core luminosities. 

Here, we briefly review the key differences that were observed in the dynamics of each model from collapse through explosion. The 15.78 \msun\ progenitor undergoes a fourth carbon shell burning phase 0.1 yr before collapse at the same time as the first oxygen shell ignition. This slows the contraction of the core and results in a silicon core mass that is smaller than that of the 15.79 \msun\ progenitor at the time of collapse. The 15.79 \msun\ progenitor undergoes its fourth carbon shell burning stage roughly 8 hr before collapse. The end result is that the 15.78 \msun\ progenitor has a sharp drop in its density profile that is not present in the 15.79 \msun\ progenitor, as shown in \citet{BrSiLe23} Figure 4. These differing density profiles lead to the 15.78 \msun\ progenitor being less compact, $\xi_{2.5}=0.136$, than the 15.79 \msun\ progenitor, $\xi_{2.5}=0.206$, where $\xi_M$ is given by Equation \eqref{eq:compact}.

Using the multiphysics \chimera\ code, these precollapse progenitors were evolved through explosion as part of the \chimera\ F-series simulations. We will denote them by their respective masses as F15.78 and F15.79. The \chimera\ code simulates CCSNe using Newtonian self-gravity with a monopole correction for general relativistic effects, spectral multigroup flux-limited diffusion four-species neutrino transport in the ray-by-ray-plus approximation, Newtonian hydrodynamics, and a nuclear reaction network. Neutrino emission and absorption interactions included are electron capture on protons and nuclei using the modern rates described in \citet{LaMaSa03} and \citet{HiMeMe03}, electron–positron annihilation and nucleon–nucleon bremsstrahlung and the corresponding inverse weak reactions. The neutrino scattering processes included are isoenergetic scattering on nuclei, and neutrino–electron and neutrino–nucleon scattering, in the latter case allowing for small-energy scattering, as described in \citet{RePrLa98,RePrLa99}. The nuclear reaction network includes a 14 species $\alpha$-chain network, with three inert species (protons, neutrons, and ${}^{56}$Fe) added actively but not reactively, as described in \citet{BrSiLe23}. The nuclear equation of state (EOS) implemented for regions in nuclear statistical equilibrium is the SFHo EOS of \citet{StHeFi13} incorporated using WeakLib \cite{Landfield_thesis}. For each model, simulations are axisymmetric with 720 non-uniform radial cells and 240 fixed angular cells of width 0.75$^\circ$. The radial extent of each model comprised the inner 3.56 \msun, reaching out to the helium shell. This corresponds to an outer radius of 61,250 km for the less compact F15.78 model and 48,960 km for the F15.79 model. The radial dimension is refined at different length scales throughout the simulation, to capture features in both the PNS and around the shock.

Each model collapses, halts collapse, and bounces back with a shock wave generated from the repulsive strong nuclear force accounted for by the SFHo EOS. The shock wave generated loses energy as it moves outwards through the supersonically infalling matter and eventually stalls, as expected from neutrino-driven CCSN theory \cite{Mull20,BuVa21,Mezz23,Janka25}. Due to the sharper density gradient in F15.78, the shock is revived 120 ms after bounce due to a drop in the ram pressure ahead of the shock while significant neutrino luminosity from accretion onto the PNS surface is still occurring. Conversely, the shock of the F15.79 model is revived approximately 220 ms after bounce after neutrinos and antineutrinos have deposited energy behind the shock, slowly building up pressure behind it until the shock overcomes the ram pressure ahead of it. Shock revival is captured in Figure 5(a) of \citet{BrSiLe23}, and explosion energy is shown in Figure 7(b). The late revival of the shock for F15.79 results in a more energetic explosion.

As the shock is revived, the explosion dynamics of each model continue to diverge. We highlight that, in particular, the mass accretion onto the PNS drops at the onset of explosion for the F15.78 model in comparison to the F15.79 model, as seen in Figure 5(c) in \citet{BrSiLe23}. This, in addition to the already less massive silicon core at the onset of collapse, results in the F15.78 model evolving a less massive PNS than the F15.79 model. Figure 8 of \citet{BrSiLe23} shows the radial and mass evolution of the PNS for both models.

\subsection{Gravitational Wave Extraction}
\label{sec:gw_extraction}
 
To extract gravitational wave strains from \chimera\ data, we utilize the equations provided in \citet{MeMaLa23}. The quadrupole moment of the transverse-traceless gravitational wave strain for an observer at time $t$ and distance $D$ from the source is given by
\begin{equation}
    h^{\mathrm{TT}}_{ij}=\frac{1}{D}\sum_{m=-2}^{+2}\frac{dN_{2m}}{dt}\left(t-\frac{D}{c}\right)f^{2m}_{ij},\label{eq:h_matter}
\end{equation}
where $i$ and $j$ span $r$, $\theta$, and $\phi$, $m$ is the azimuthal component of the spherical harmonics, $f^{2m}_{ij}$ are the tensor spherical harmonics, and $N_{2m}$ is defined as
\begin{equation}
    N_{2m}\equiv\frac{G}{c^4}\frac{dI_{2m}}{dt}.
\end{equation}
Taking the time derivative of the quadrupole moment $I_{2m}$ and using the continuity equation for the density yields
\begin{align}
    &N_{2m}=\frac{16\sqrt{3}\pi G}{15 c^4}\int^{2\pi}_{0}d\varphi'\int^\pi_0d\vartheta'\int^b_adr'r'^3\nonumber\\
           &\times\left[2\rho v^rY^*_{2m}\sin{\vartheta'}+\rho v^\vartheta\sin{\vartheta'}\frac{\partial}{\partial\varphi'}Y^*_{2m}+\rho v^\varphi\frac{\partial}{\partial \varphi'}Y^*_{2m}\right]\nonumber\\
           &-\frac{16\sqrt{3}\pi G}{15 c^4}\int_0^{2\pi}d\varphi'\int_0^\pi d\vartheta'Y^*_{2m}\sin{\vartheta'}\left(r^4_b\rho_bv_b^r-r^4_a\rho_av^r_a\right)\label{eq:N2m}
\end{align}
with $v^i$ being the linear fluid velocity and the final term being a surface contribution from the radial shell with inner radius $r_a$ and outer radius $r_b$. In the two-dimensional models considered in this work, the only nonzero gravitational wave strain is given by
\begin{equation}
    h_+=\frac{h^{\mathrm{TT}}_{\theta\theta}}{r^2}.
\end{equation}

\begin{figure*}
    \centering
    \includegraphics[width=\columnwidth]{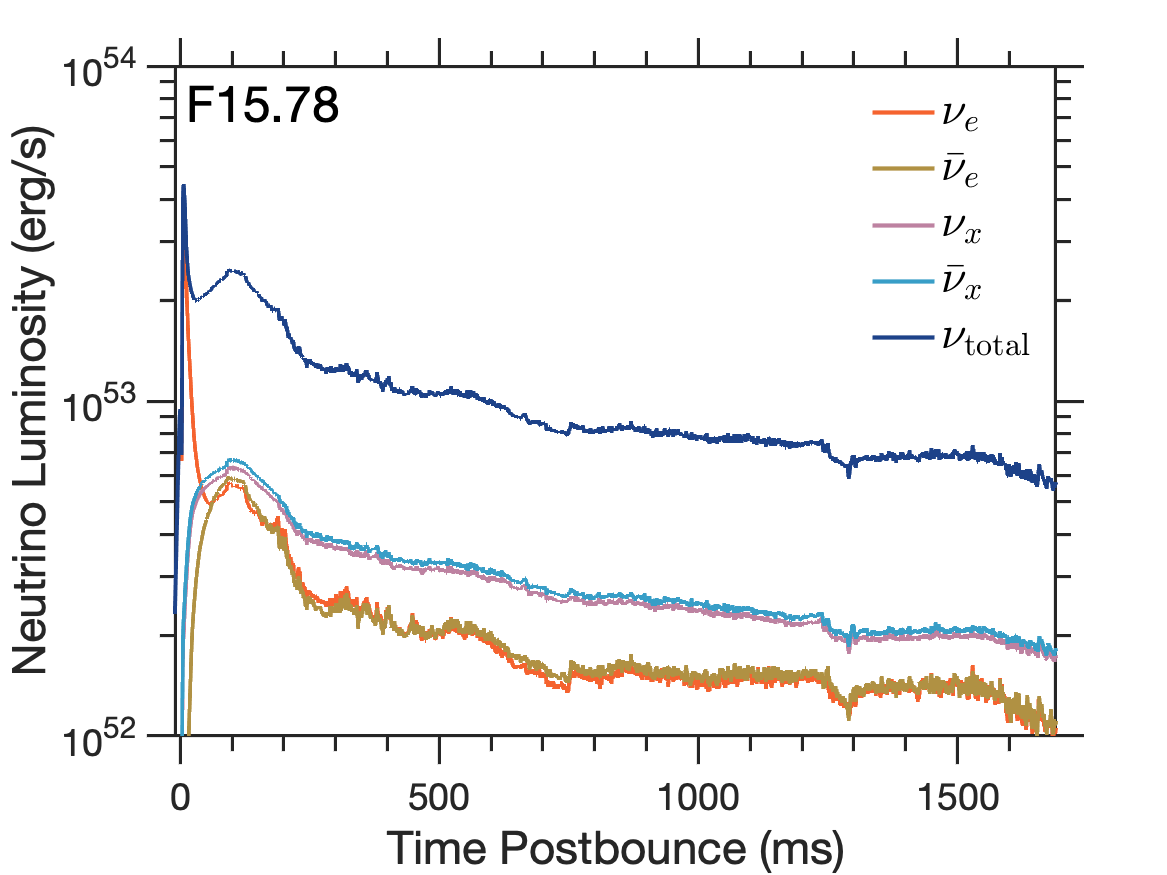}
    \includegraphics[width=\columnwidth]{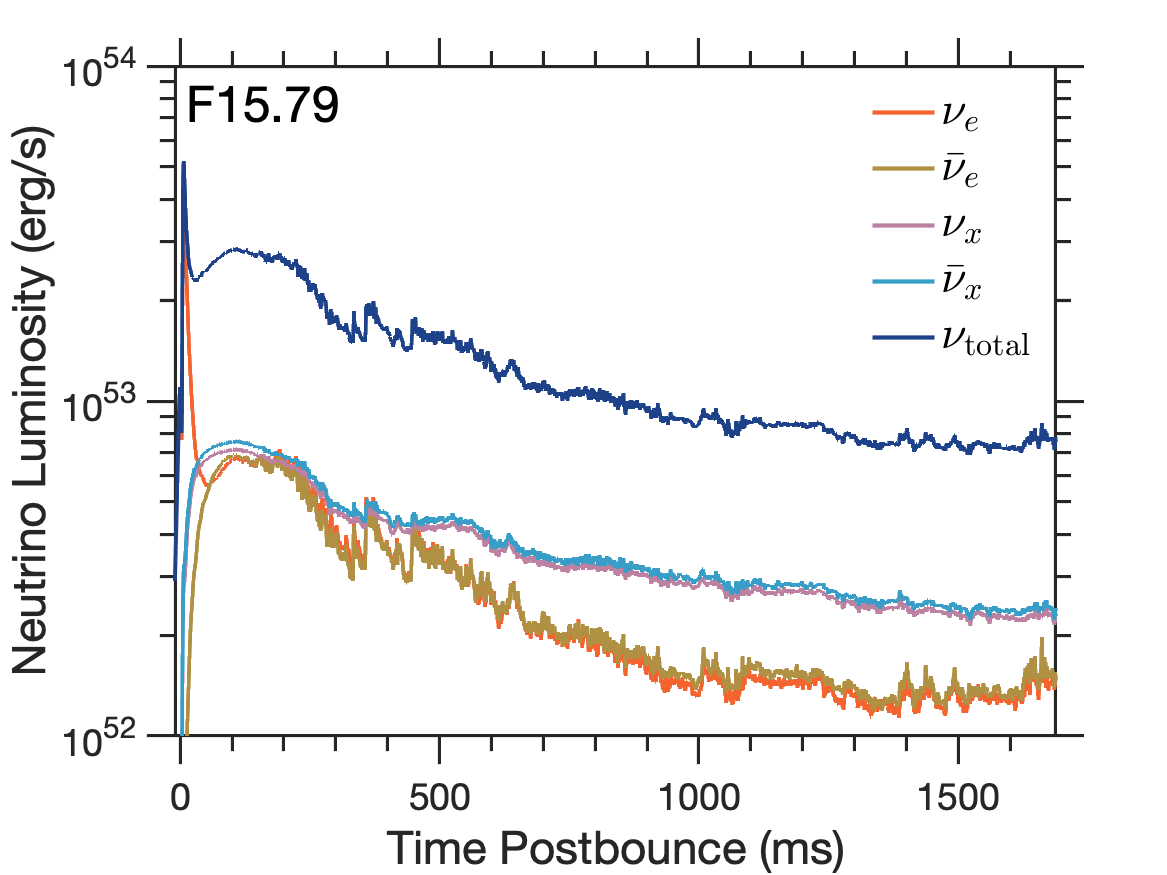}\hfill
    \includegraphics[width=\columnwidth]{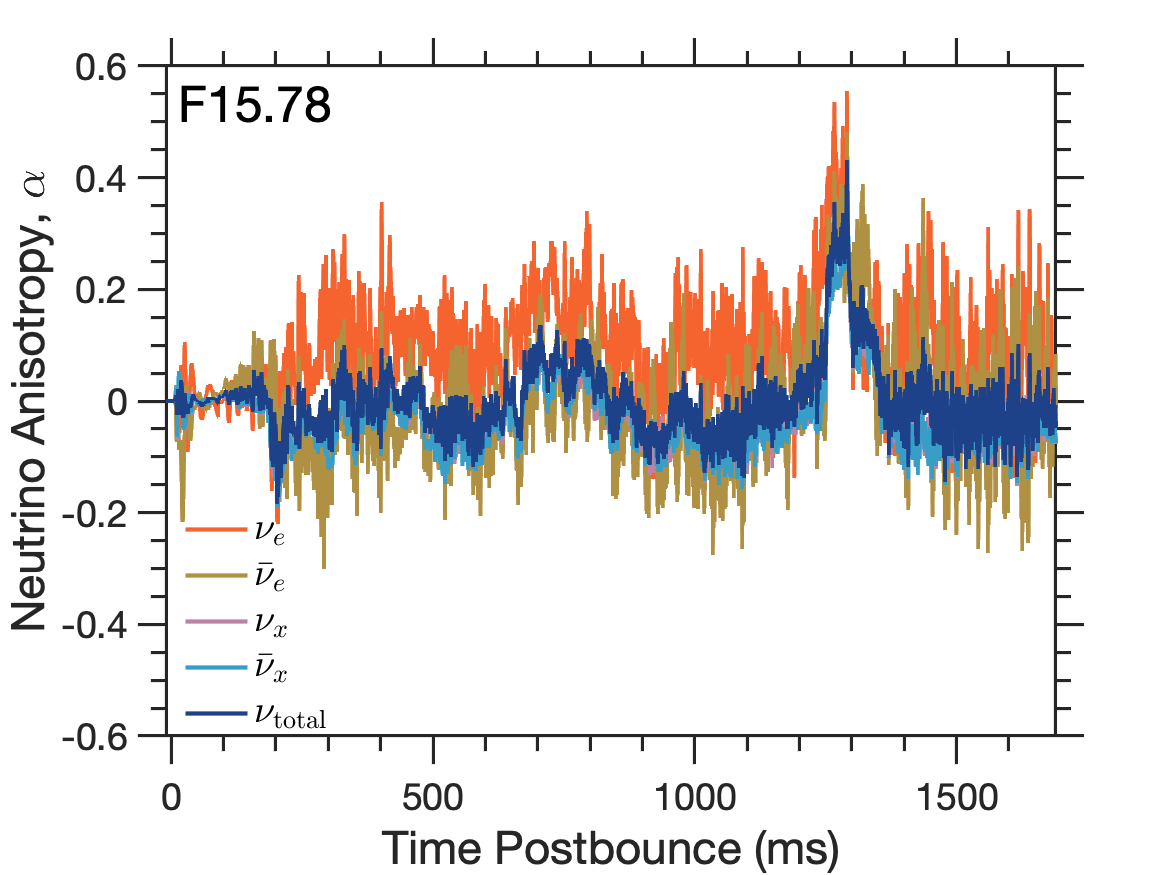}
    \includegraphics[width=\columnwidth]{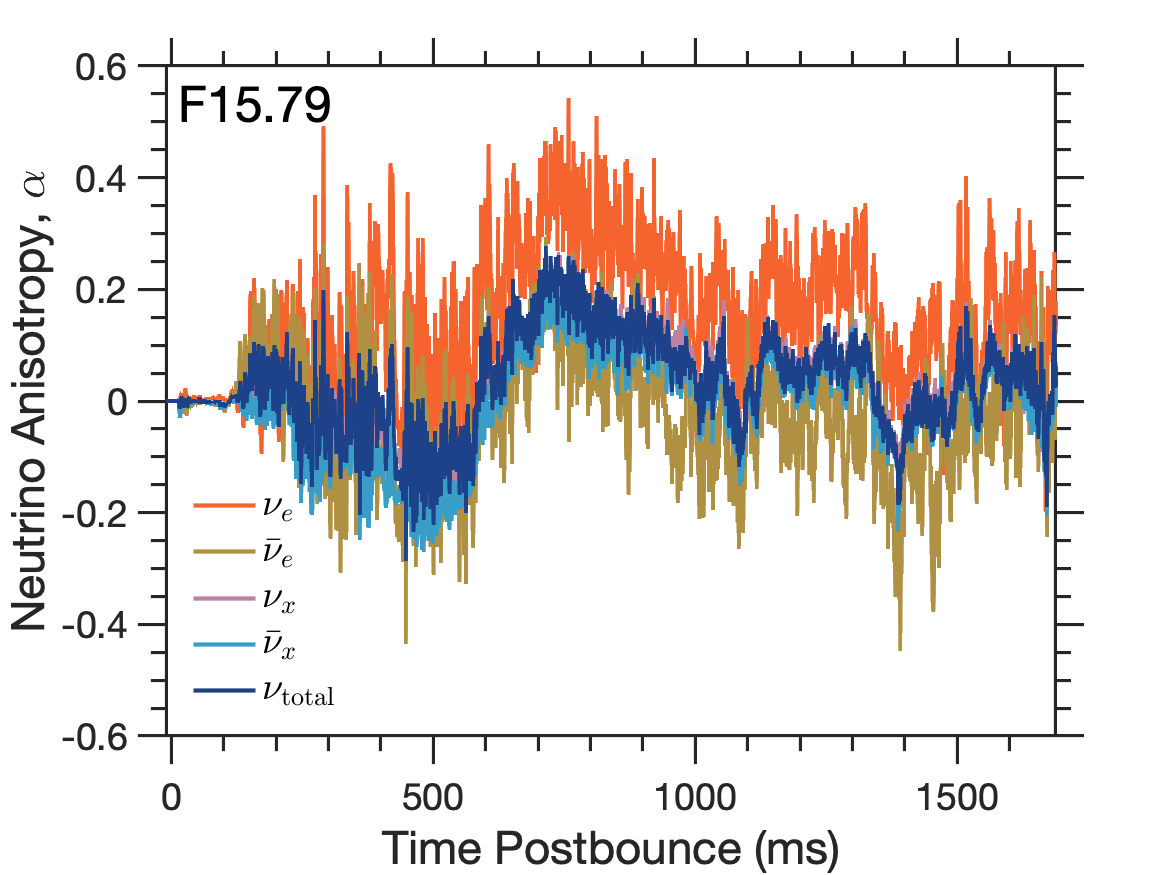}
    \caption{Angle-integrated energy luminosity of neutrinos passing through a sphere of radius 500 km (top) and anisotropy parameter (bottom) for each model. For each quantity, neutrino species are denoted by color, with orange for electron neutrinos, yellow for anti-electron neutrinos, purple for heavy neutrinos ($\mu+\tau$ neutrinos), and light blue for anti-heavy neutrinos. Dark blue indicates the total luminosity/anisotropy from all neutrino species.}
    \label{fig:neutrino_luminosity}
\end{figure*}

In  previous studies \cite{MeMaLa20,MeMaLa23,MuMeLe25pre}, in addition to the total gravitational wave strain, the computational domain was decomposed into five regions within which the gravitational wave strains could be computed. These regions were, in order moving outward, the convective region of the proto-neutron star (PNS), the convective overshoot layer of the PNS, the surface layer of the PNS defined from the $10^{12}$ to the $10^{11}$ g cm$^{-3}$ density contours, the net cooling layer behind the shock, and the net heating layer behind the shock (\citet{MuMeLe25pre} extended this last region out to the computational grid boundary instead of the shock radius). For the three-dimensional models considered in those studies, each of these regions was well defined. In the two-dimensional models we consider here, this is not the case, as the low plume counts in the convective regions make the boundaries based on means unsteady and not useful. We have therefore omitted a regional analysis from this study.


At late times during the CCSN simulation, refinement of the radial grid around the shock happens frequently as it moves through the computational domain. For the calculation of gravitational wave strains, this can result in the appearance of sudden changes in velocity at cell interfaces that are purely numerical. From one time step to the next, the fluid moves with a physical velocity between cells. However, as the grid refines, the boundaries of these cells can change on the order of kilometers between time steps. This can cause the gravitational wave computation to calculate non-physical velocity derivatives for boundaries that moved, which results in non-physical ``combs'' to appear in the gravitational wave strains. Since this effect is localized around the shock, it does not affect the overall strain computation, only the contribution from the cells around the shock. Once the shock is revived and explosion begins, the region far beyond the PNS surface is dominated by the evolution toward the gravitational wave memory, which exists in the 10s of Hz range of the gravitational wave signal \cite{MeMaLa20, MeMaLa23, MuMeLe25pre}. As the high-frequency ``combs'' are non-physical, we implement a low-pass filter, attenuated at 100 Hz, on the signal produced at radii larger than the $10^{10}$ g cm$^{-3}$ spherically averaged density contour (i.e., $r\ge R|_{\rho=10^{10}}$). This filtering retains the physical, low-frequency component of the gravitational wave strains that comes from outside the PNS, and it also retains the physical high-frequency component of the gravitational wave strains that comes from regions deep within and just outside of the PNS, as shown in \citet{MeMaLa23} and \citet{MuMeLe25pre}. The low-pass filter is applied at 933 ms and 624 ms postbounce for F15.78 and F15.79, respectively.

In addition to the gravitational wave signal from the matter motion, we also present the gravitational waveforms from the anisotropic emission of neutrinos. We extract the waveforms following the methods discussed in \citet{RiMeMu25pre}, using
\begin{equation}
    h^\nu_{+}(t, \alpha', \beta') = \frac{2}{D} \int_{0}^{t} dt' L_{E}^{\nu}(t') \alpha^\nu_{+}(t', \alpha', \beta'),\label{eq:nu_gw}
\end{equation}
where $L_{E}^{\nu}$ is the global, angle integrated energy luminosity of neutrinos and $\alpha^\nu_{+}$ is the anisotropy parameter introduced in \citet{MuJa97}. Following the notation in \citet{RiMeMu25pre}, angular coordinates with a $'$ superscript correspond to the coordinates of a particular radiation solid angle in the ray-by-ray approximation used for neutrino transport in \chimera\ as defined in the source frame, while coordinates without superscripts correspond to coordinates in the observer's frame. Additionally, the angular coordinates of the observer in the source frame are denoted by $(\alpha',\beta')$. In Equation \eqref{eq:nu_gw}, $L_{E}^{\nu}$ is given by
\begin{equation}
L_{E}^{\nu}=\int_{4\pi}d\Omega'\frac{d L^\nu}{d\Omega'}\left(\Omega',t'\right), \label{eq:L_E}
\end{equation}
and $\alpha^\nu_{+}$ is defined as
\begin{equation}
    \alpha_+(t,\theta,\phi)=\frac{1}{L_{E}^\nu}\int_{4\pi}d\Omega'W_+(\theta,\phi,\Omega')\frac{dL_E^\nu\left(\Omega',t'\right)}{d\Omega'},\label{eq:ap}
\end{equation}
where $W_+$ is an angular weight defined in the source frame by Equation 32 in \citet{RiMeMu25pre} as
\begin{equation}
    W_+(\alpha',\beta',\theta',\phi')=\left(1+\frac{z'}{r'}\right)\frac{x'^2-y'^2}{x'^2+y'^2}
\end{equation}
and $\frac{dL_E^\nu}{d\Omega'}$ is the differential neutrino luminosity computed as
\begin{equation}
    \frac{dL_E^\nu}{d\Omega'}=\frac{4\pi r'^2}{\alpha^4}\int dE\psi^{(1)}(\phi',\theta',r',\nu,E)E^3
\end{equation}
with $\alpha$ as the lapse function, $E$ as the neutrino energy, and $\psi^{(1)}$ as the first angular moment of the neutrino distribution computed in the Eulerian frame of reference. Note the use of natural units, $G=c=h=1$, so that we exclude the additional factor of $h^{-3}c^{-2}$. The differential neutrino luminosity is extracted at $r'=500$ km.

In Figure \ref{fig:neutrino_luminosity}, we plot the neutrino energy luminosity $L^\nu_E$ computed using Equation \eqref{eq:L_E} in the top panels and the anisotropy parameter computed using Equation \eqref{eq:ap} in the bottom panels, for both progenitor models. The quantities for each neutrino species are shown in different colors, with the total quantity  (i.e., the sum of each species contribution) shown in dark blue. For Equation \eqref{eq:nu_gw}, the total energy luminosity and total anisotropy parameter are used.

\section{Results}
\label{sec:results}

\subsection{Gravitational Wave Strains}
\label{sec:gw_strains}

\begin{figure}
\centering
  \includegraphics[width=\columnwidth]{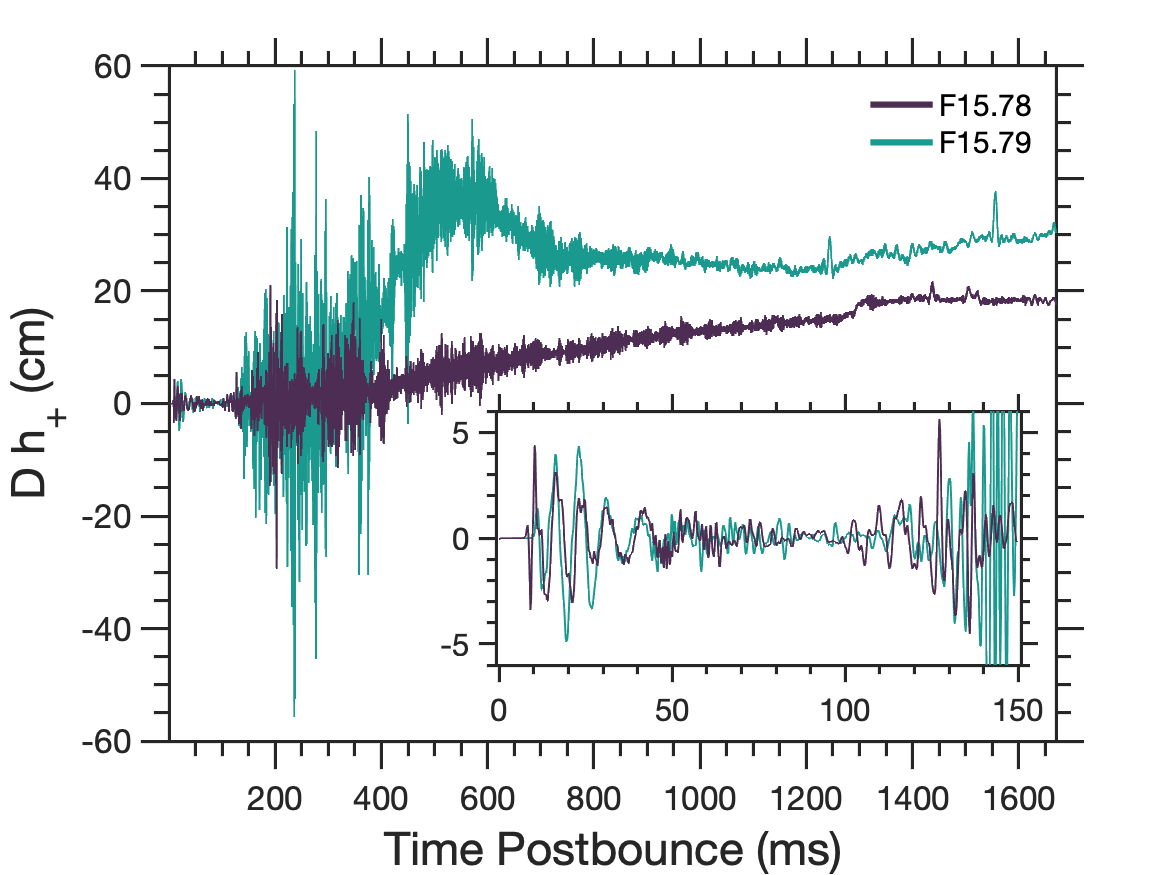}
  \caption{$h_+$-polarized strains multiplied by a distance of 10 kpc as viewed from the $z$-axis of the simulation for F15.78 (purple) and F15.79 (teal). The strain for the first 150 ms after bounce is shown, with a modified vertical axis, in the inset plot.}
  \label{fig:strain_total}
\end{figure}

\begin{figure*}[p]
\centering
  \includegraphics[width=\columnwidth]{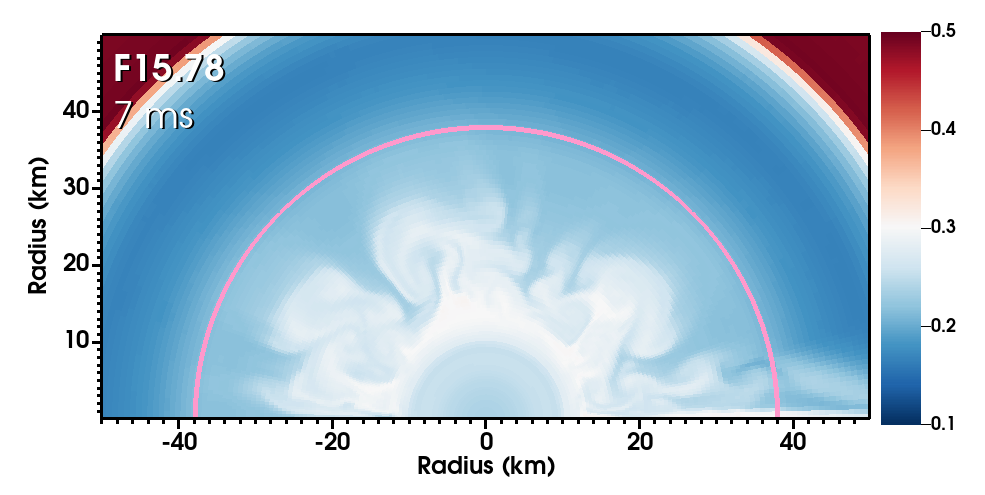}
  \includegraphics[width=\columnwidth]{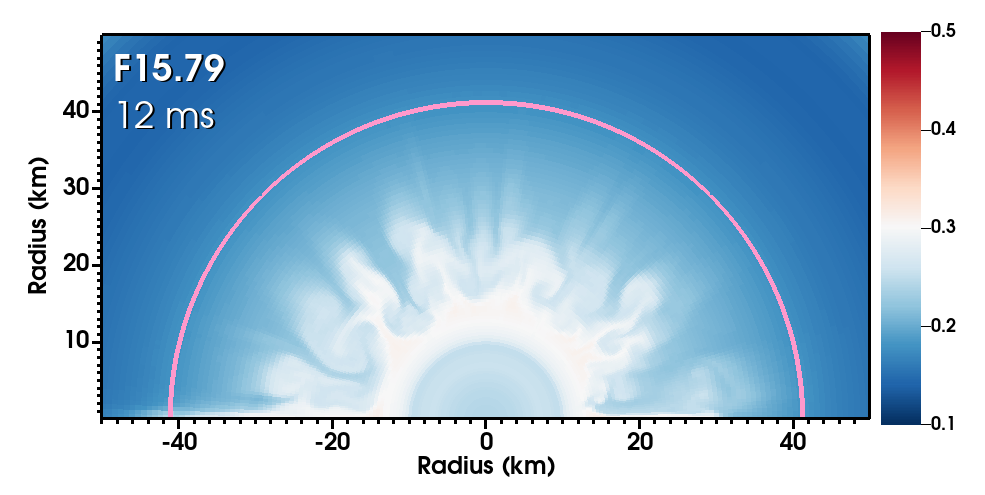}
  \hfill
  \includegraphics[width=\columnwidth]{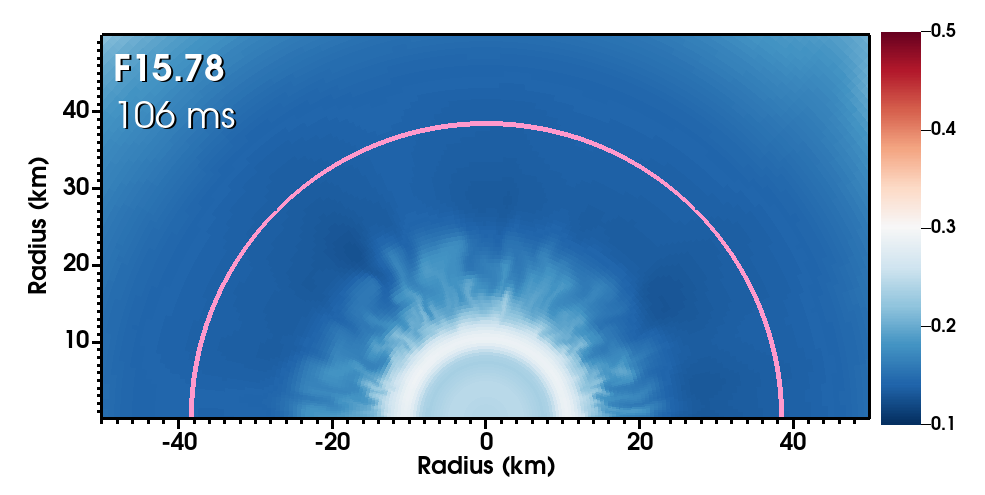}
  \includegraphics[width=\columnwidth]{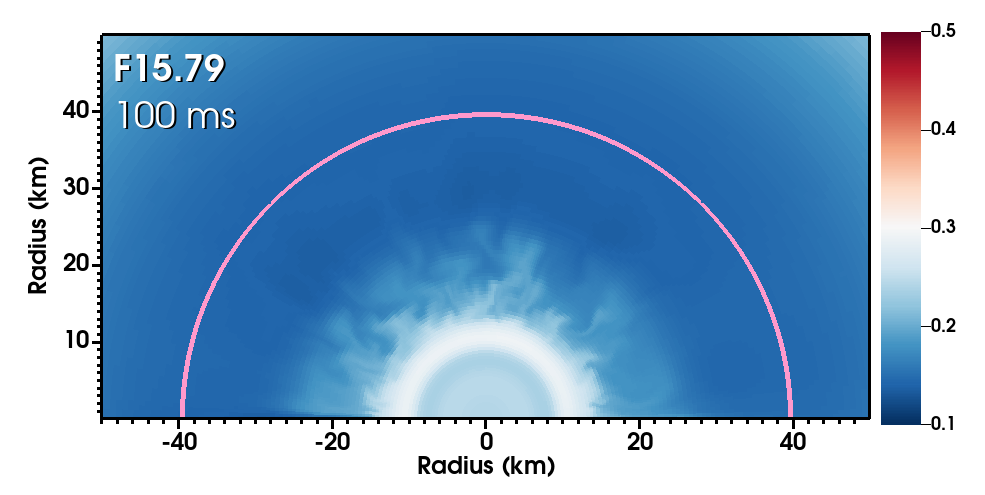}
  \hfill
  \caption{Plot of electron fraction within the inner 50 km of the star for F15.78 (left) and F15.79 (right). The top panels are shown at 7 ms for F15.78 and 12 ms for F15.79. The bottom panels are shown at 106 ms for F15.78 and 100 ms for F15.79. The color axis represents electron fraction, and the pink contour corresponds to the outer boundary of Region I.}
  \label{fig:pc_Ye}
\end{figure*}

Figure \ref{fig:strain_total} shows the gravitational wave strains from matter multiplied by a distance of 10 kpc as viewed from the $z$-axis of the simulation, computed using Equation \eqref{eq:h_matter} for each model. The inset of Figure \ref{fig:strain_total} shows the gravitational wave signal immediately after bounce. Within $\sim$10 ms, both models experience prompt convection driven by lepton and entropy gradients deep within the PNS. In the top panel of Figure \ref{fig:pc_Ye}, we show the electron fraction for each model at the onset of prompt convection. The low-frequency component of the gravitational wave signal at early times is similar for both models, with the most prominent frequency being between 120--140 Hz from 10--50 ms after bounce. The impact of prompt convection on the gravitational wave signal has settled by approximately 50 ms after bounce, as seen in the inset of Figure \ref{fig:strain_total}, and each model has a short quiescent period with weak gravitational wave generation for approximately 50 ms. By $\sim$100 ms after bounce, each model develops Ledoux convection deep within the PNS as shown in the bottom panel in Figure \ref{fig:pc_Ye}. 

 \begin{figure*}[p]
\centering
  \includegraphics[width=\columnwidth]{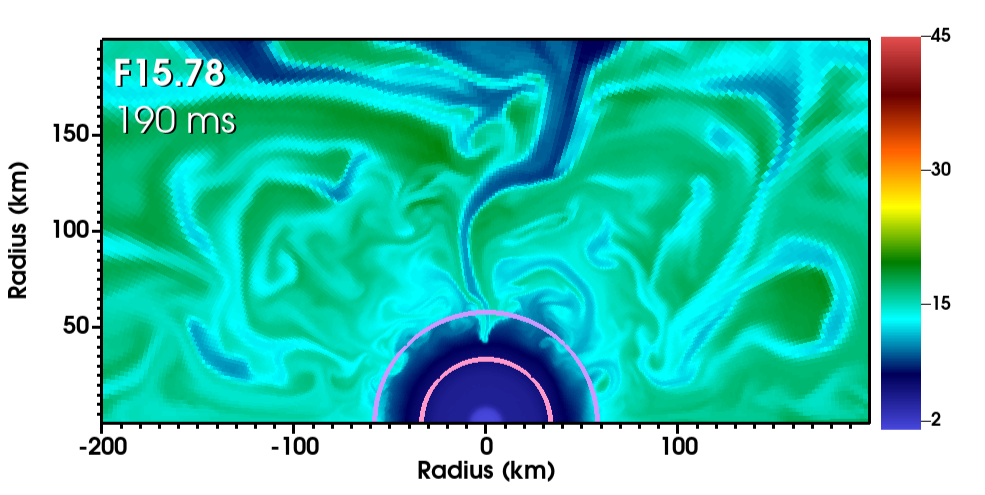}
  \includegraphics[width=\columnwidth]{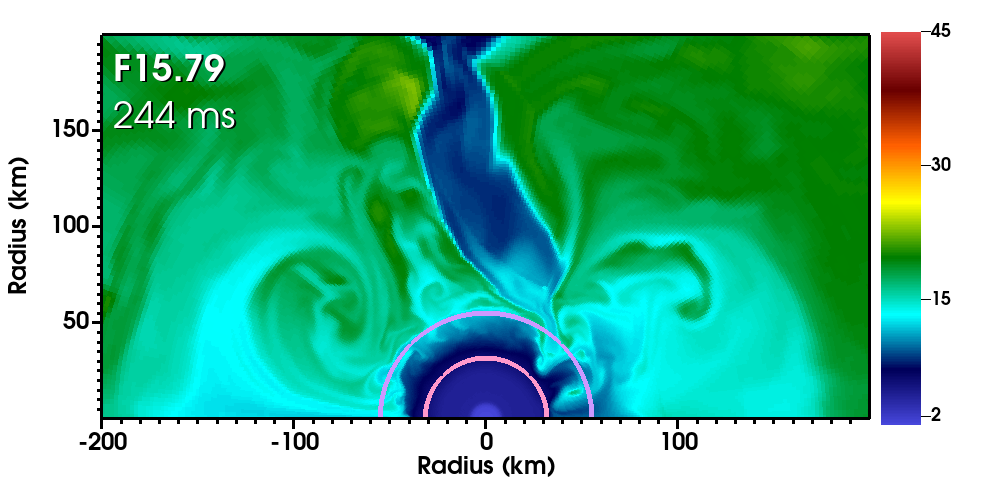}
  \includegraphics[width=\columnwidth]{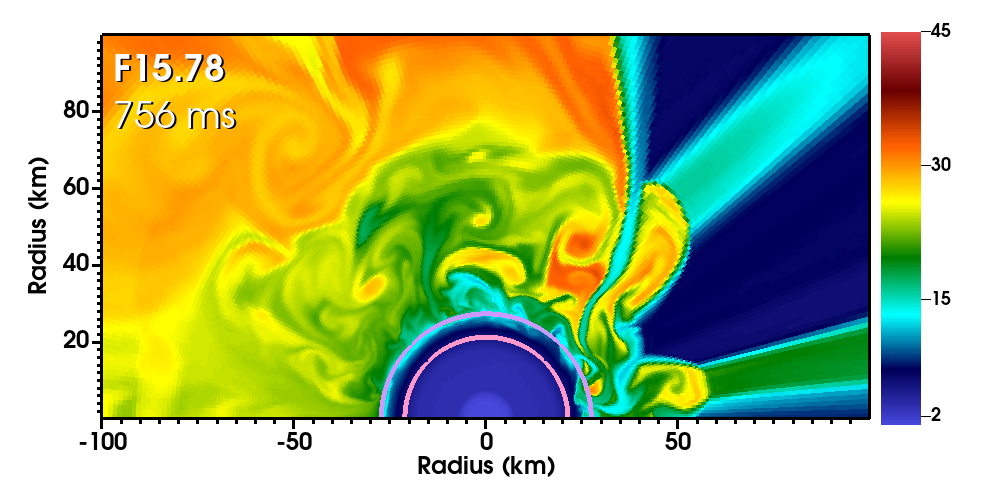}
  \includegraphics[width=\columnwidth]{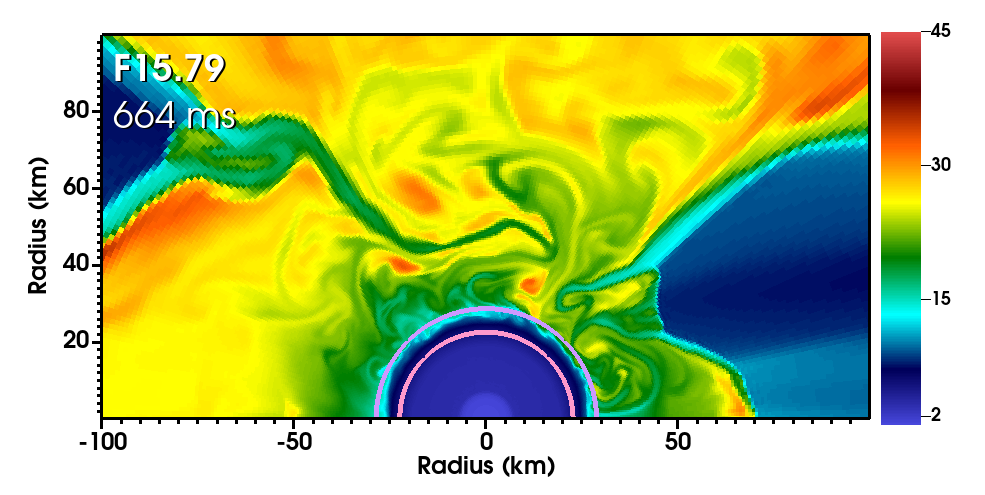}
  \caption{Specific entropy in and around the PNS for F15.78 (left) and F15.79 (right). The purple contour denotes the  
  $10^{10}$ g cm$^{-3}$ spherically-averaged density contour (i.e., outside the PNS) and the pink contour denotes the $10^{12}$ g cm$^{-3}$ spherically-averaged density contour (i.e., deep inside the PNS).Top panels show plots at 190 ms and 244 ms after bounce for F15.78 and F15.79, respectively. Bottom panels show plots at 756 ms and 664 ms after bounce for F15.78 and F15.79, respectively.}
  \label{fig:78_ent}
\end{figure*}

\begin{figure}
    \centering
    \includegraphics[width=\columnwidth]{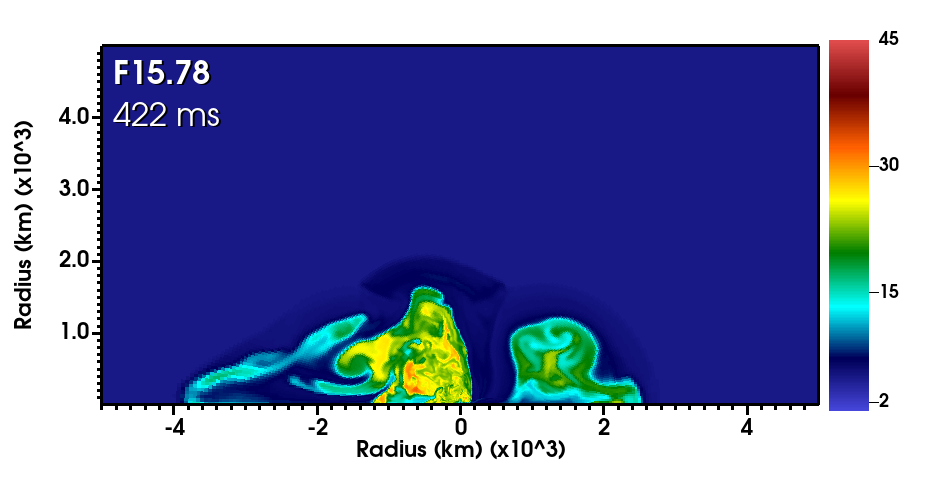}
    \includegraphics[width=\columnwidth]{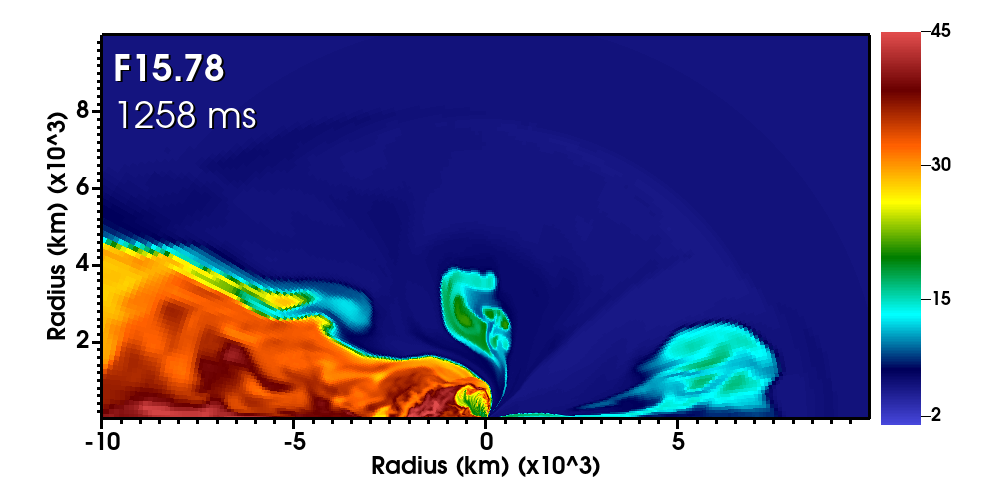}\hfill
    \caption{Entropy snapshots for F15.78 at 442 ms and 1258 ms, respectively, displaying key hydrodynamic features that contribute to the evolution of the memory resulting from the matter outflows.}
    \label{fig:F78_Mem_ent}
\end{figure}

After $\sim$140 ms after bounce, the gravitational wave strains in the two models begin to diverge, and the magnitude of the gravitational wave strains for F15.79 are generally larger than the strains for F15.78. The only exception is over a duration $\sim$15 ms beginning at $\sim$190 ms after bounce for F15.78, where the gravitational wave strains are comparable to those of F15.79. For each model, the period with the greatest magnitude of gravitational wave strain is between $\sim$150--700 ms after bounce, neglecting the positive strain contribution from the ramp up to gravitational wave memory. This coincides with the times with the greatest neutrino accretion luminosities (i.e., the luminosities of neutrinos produced through accretion of matter onto the PNS) as shown in Figure 16 of \citet{BrSiLe23}. Which is not surprising considering the relative importance off accretion onto the PNS in sourcing high-frequency gravitational waves in two-dimensional models. However, we note that this study does not directly examine the source of the gravitational wave signal.

Nonetheless, it remains possible to connect specific hydrodynamical events to the gravitational wave signal. In Figure \ref{fig:78_ent}, the left panels show the specific entropy in the region near the PNS for F15.78 at 190 ms after bounce (top) and 756 ms after bounce (bottom). In the top panel, a low-entropy accretion funnel is seen (dark blue) penetrating the $10^{10}$ g cm$^{-3}$ spherically-averaged density contour (purple), almost reaching the inner region of the PNS defined (see \citet{MuMeLe25pre}) at the $10^{12}$ g cm$^{-3}$ density contour (pink). This corresponds precisely with the time at which there is a large spike in the gravitational wave strain for F15.78, in Figure \ref{fig:strain_total}. In the bottom panel, we see the presence of a large accretion funnel. However, it does not penetrate the $10^{10}$ g cm$^{-3}$ density contour. Thus, the corresponding spike in the gravitational wave strain is much smaller at $\sim$750 ms after bounce in Figure \ref{fig:strain_total}. 

The right panels panel of Figure \ref{fig:78_ent} show the entropy plots at 244 ms (top) and 664 ms (bottom) after bounce. For F15.79, the largest gravitational wave strain in Figure \ref{sec:gw_strains} occurs $\sim$240 ms after bounce, which corresponds to a large, low-entropy accretion funnel penetrating almost all the way to the inner PNS at 244 ms in Figure \ref{fig:78_ent}. In the bottom-right panel of Figure \ref{fig:78_ent}, we show that for F15.79 the accretion funnels penetrate less 664 ms after bounce than at earlier times, corresponding to smaller gravitational wave strains. For both models, these smaller accretion funnels that rarely penetrate past the $10^{10}$ g cm$^{-3}$ density contour after $\sim$640 ms after bounce agree with the much lower accretion luminosities seen at these times in Figure 16 of \citet{BrSiLe23}.

\begin{figure}
    \centering
    \includegraphics[width=\columnwidth]{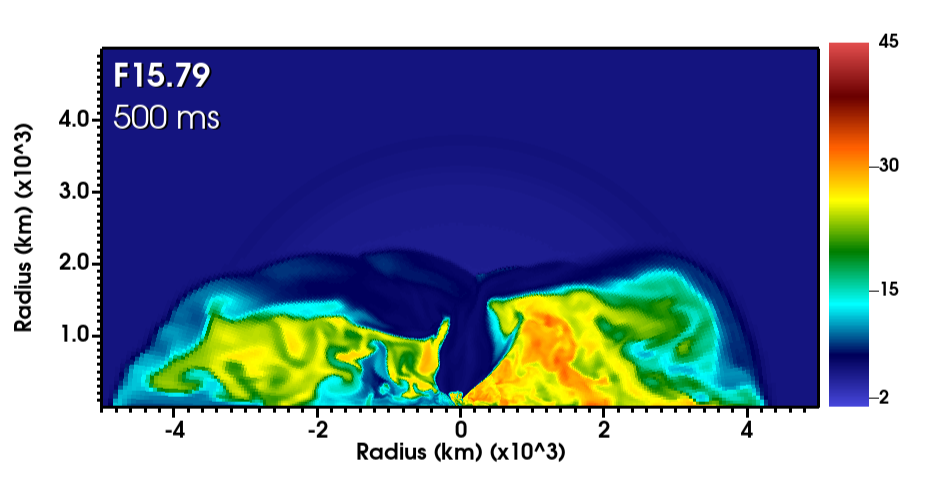}
    \includegraphics[width=\columnwidth]{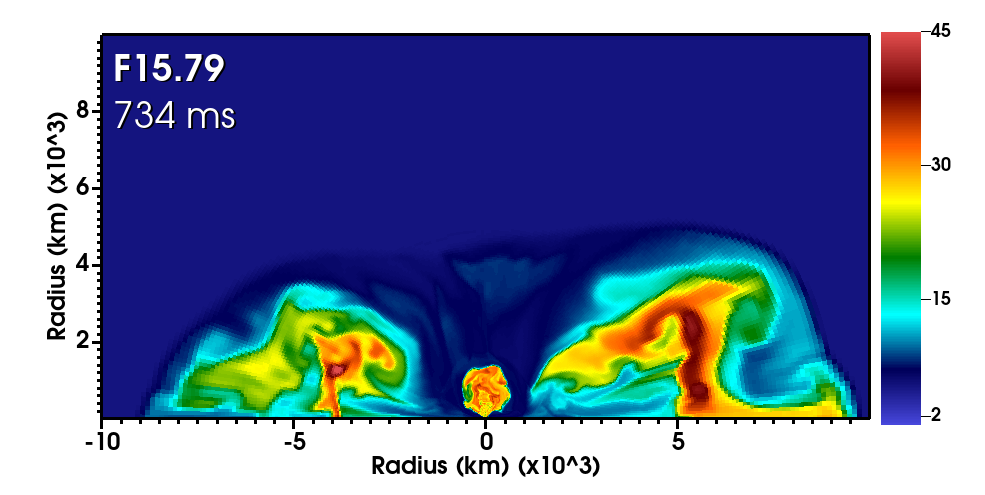}
    \includegraphics[width=\columnwidth]{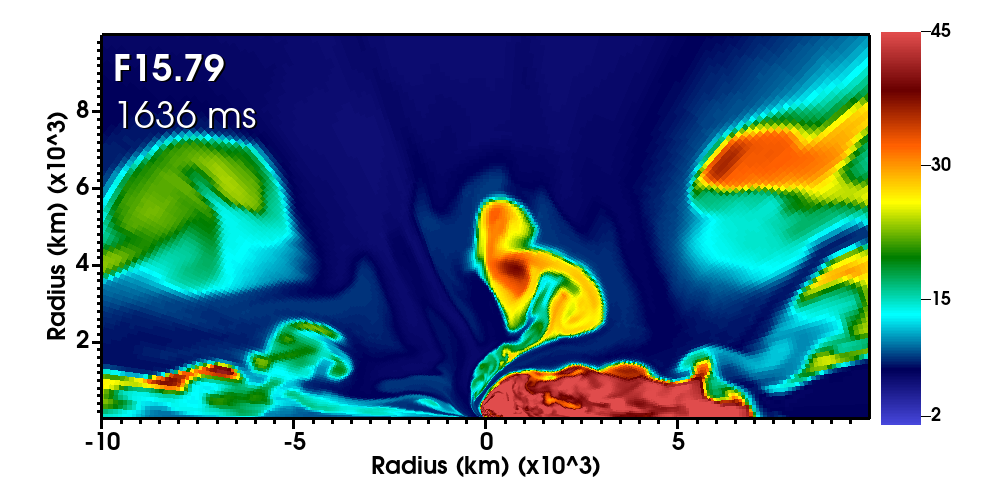}
    \caption{Entropy snapshots for F15.79 at 500 ms, 734 ms, and 1636 ms, respectively, displaying key hydrodynamic features that contribute to the evolution of the memory resulting from the matter outflows.}
    \label{fig:F79_Mem_ent}
\end{figure}

The high-frequency component of the gravitational wave strain is comparatively weak for both models $\sim$700 ms after bounce and beyond. At this time, the more prominent feature of the signal is the ramp up to gravitaitonal wave memory. The gravitational wave memory is the sum of contributions from gravitational waves emitted due to explosion and to the anisotropic emission of neutrinos. Regarding the former, we present in Figures \ref{fig:F78_Mem_ent} and \ref{fig:F79_Mem_ent} snapshots of the entropy for each model at different times. For both models, after the shock is revived the prolate shock expansion causes a low-frequency increase in gravitational wave strain seen as a ``tail" in Figure \ref{fig:strain_total}. Bilateral entropy plumes signifying matter moving outward at 422 and 500 ms after bounce for each model, respectively, are evident. Interestingly, in the F15.79 model we observe a plume generated perpendicular to the axis of explosion between 600--800 ms after bounce, shown at 734 ms after bounce in Figure \ref{fig:F79_Mem_ent}. This additional plume decreases the quadrupole moment of the explosion, and we see this effect in an overall decrease of the gravitational wave strain as well. No such feature develops in F15.78. Near the end of the simulation, we see that both models exhibit plumes moving outward along a single direction.

\begin{figure}
    \centering
    \includegraphics[width=\columnwidth]{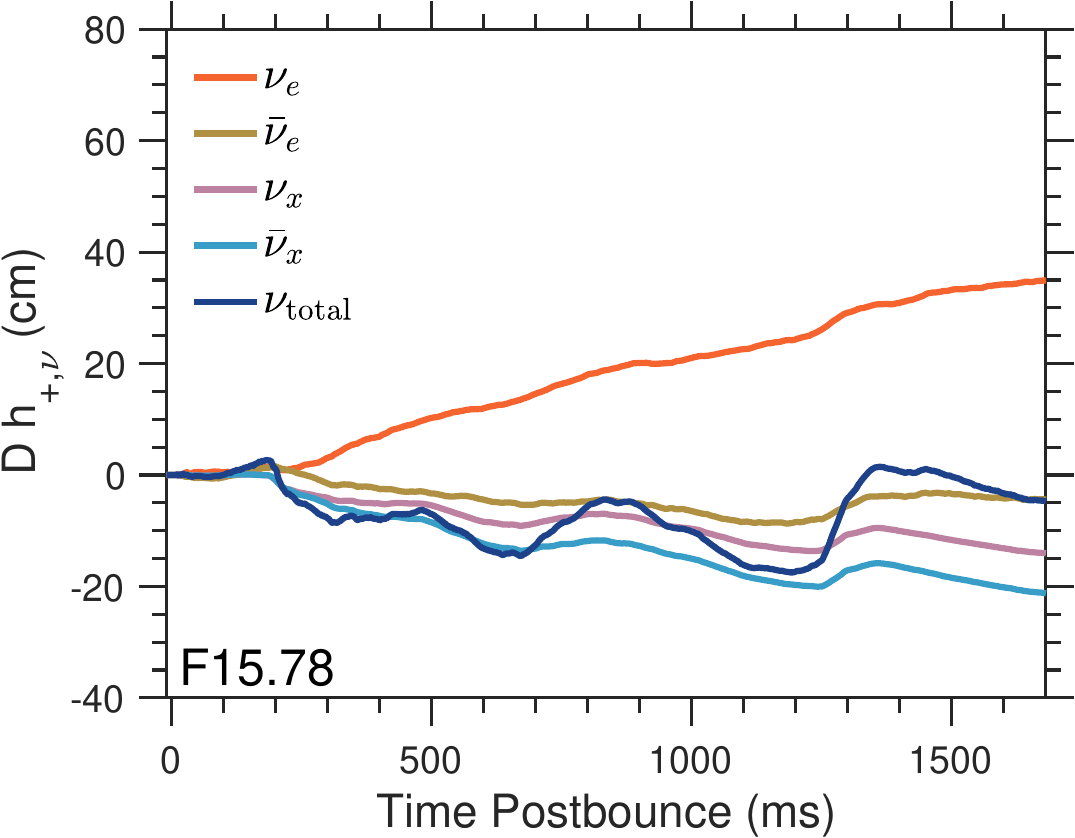}
    \includegraphics[width=\columnwidth]{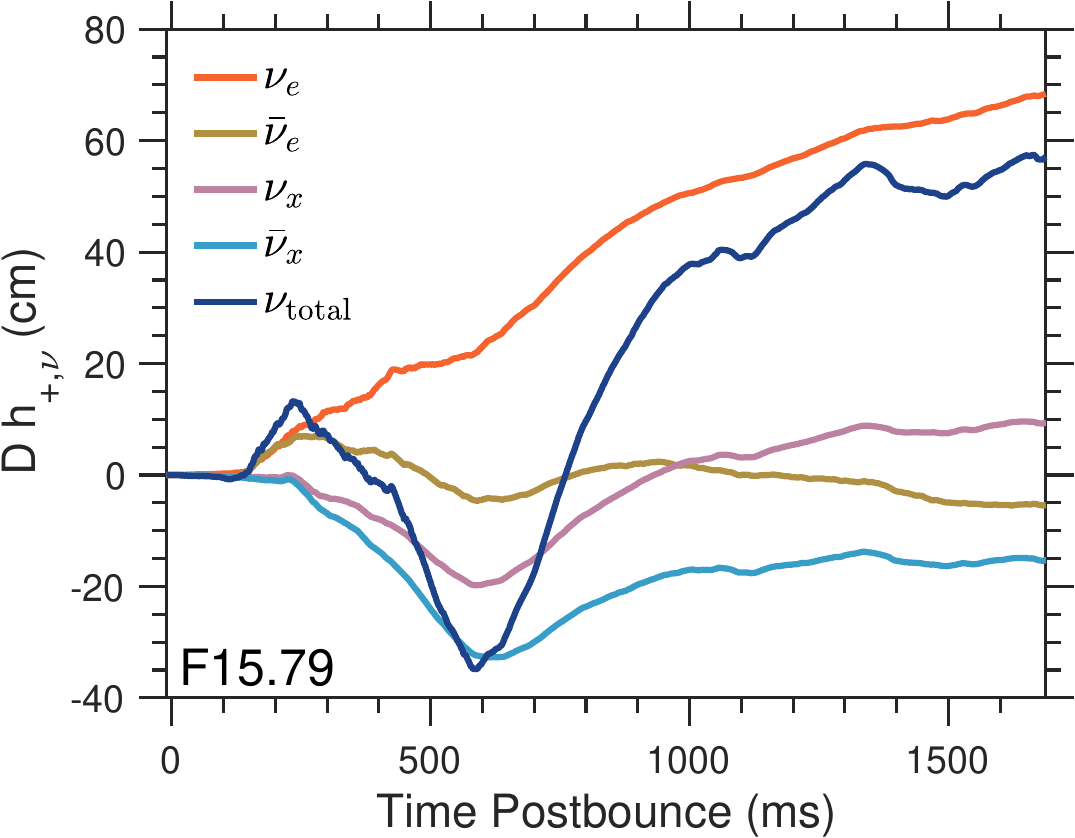}
    \caption{Gravitational wave strain from the anisotropic emission of neutrinos, multiplied by a distance of 10 kpc.}
    \label{fig:neutrino_strain}
\end{figure}

\begin{figure}
    \centering
    \includegraphics[width=\columnwidth]{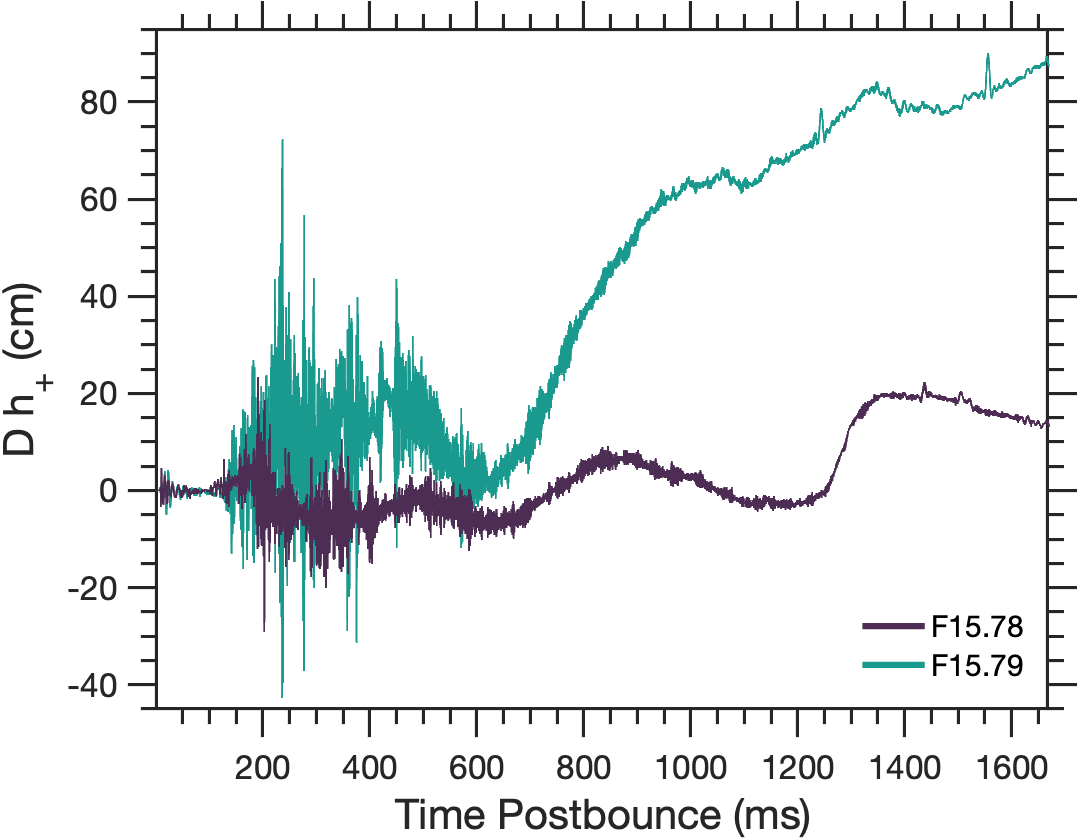}
    \caption{Total $h_+$-polarized gravitational wave strain from both the anisotropic emission of neutrinos and matter outflow, multiplied by a distance of 10 kpc for F15.78 (purple) and F15.79 (teal). The observer of the explosion is aligned along the $z$-axis of the simulation.}
    \label{fig:matter+neutrino_strain}
\end{figure}

In Figure \ref{fig:neutrino_strain}, we present the neutrino component of the gravitational wave signal, for both models, using the methods documented in \citet{RiMeMu25pre}. We plot the contribution to the gravitational wave strain from each neutrino species, as well as the total. The imprint of increased neutrino emission due to prolonged accretion of matter onto the PNS in F15.79 is seen in the larger neutrino-induced gravitational wave strains. The detected gravitational wave strain would of course be the sum of the contributions from matter motion and neutrino emission, shown in Figure \ref{fig:matter+neutrino_strain}. Comparing this figure with Figure \ref{fig:strain_total}, we see that the addition of the contribution from neutrino emission has two significant effects: (1) There is a significant change in the amplitude of the gravitational wave memory. (2) A low-frequency modulation is introduced in the evolution to the memory beginning $\sim$150--200 ms after bounce.  

\subsection{Gravitational Wave Spectrograms}
\begin{figure}
    \centering
    \includegraphics[width=\columnwidth]{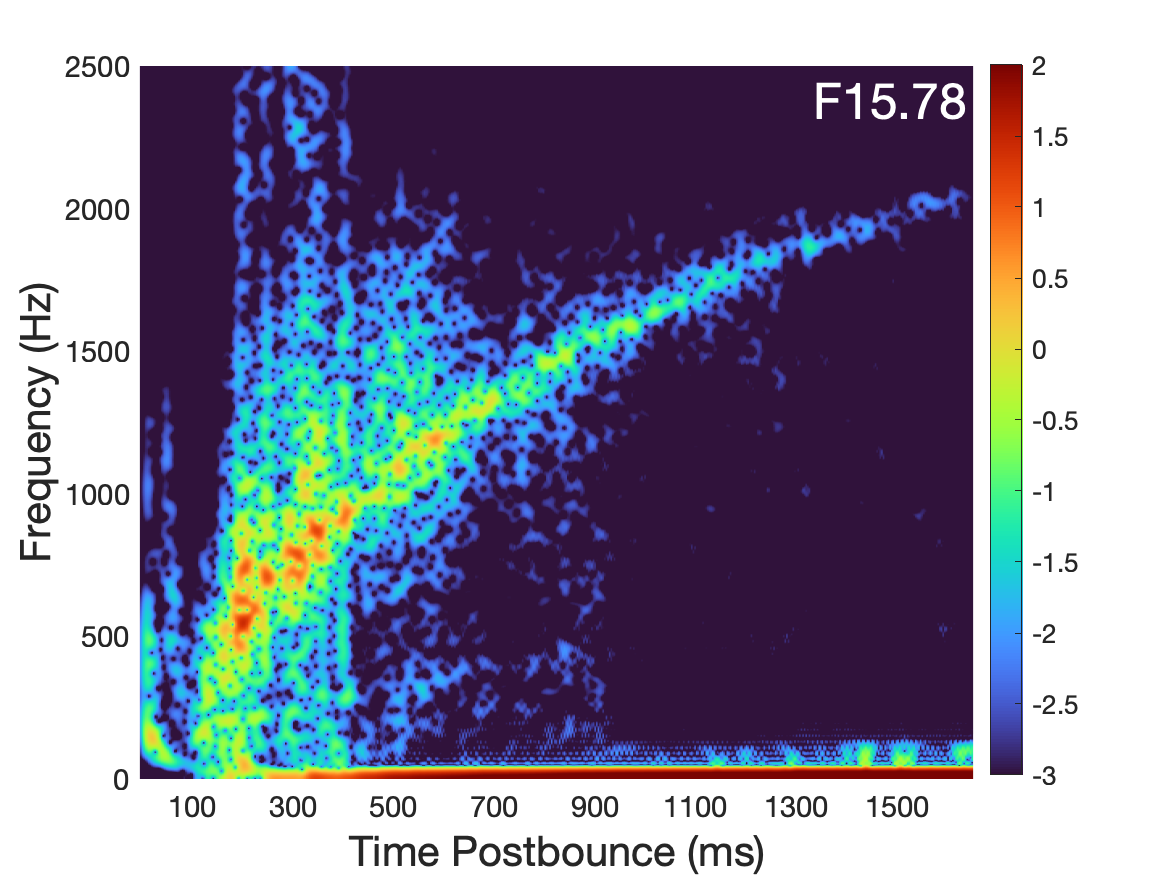}
    \includegraphics[width=\columnwidth]{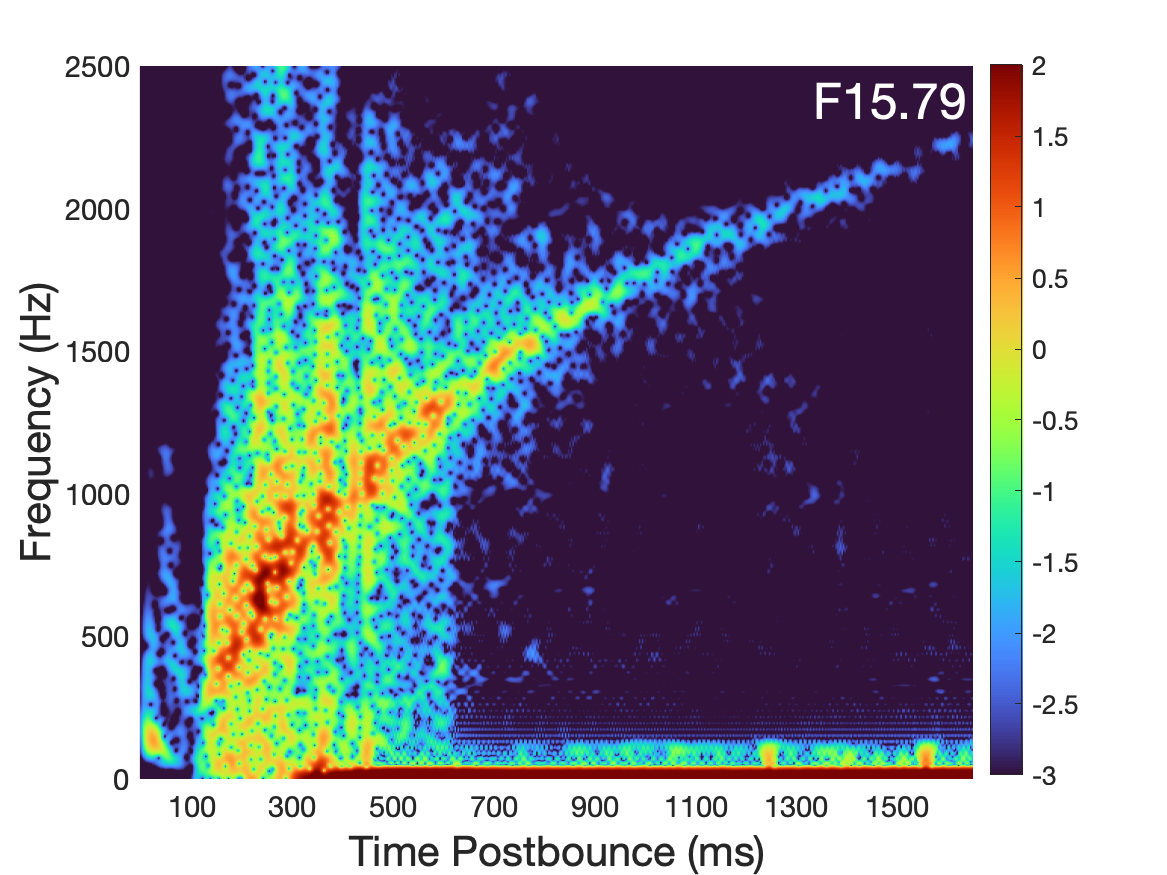}
    \caption{Spectrograms for F15.78 and F15.79, with the color axis depicting the logarithm of the power spectral density.}
    \label{fig:Spec_total}
\end{figure}

We present spectrograms of the gravitational wave signal for each model, in Figure \ref{fig:Spec_total}, computed using the methods described in \citet{MuCaMe24}. Our Kaiser windowing scheme has a leakage parameter of 0.85, a time window of 45 ms, and an overlap percentage for each time window of 93.3\%. The color axis for each spectrogram shows the base-10 logarithm of the power spectral density. The F15.78 spectrogram not only has a weaker overall signal, it also shows a narrower band structure. This reflects the fewer and less energetic accretion events onto the PNS in this model, which is in accordance with Figure 5(c) of \citet{BrSiLe23}. Likewise, the low-frequency component associated with turbulent neutrino-driven convection and SASI motion is noticeably weaker for F15.78. This is due to the relatively prompt shock revival that does not allow for extensive neutrino heating and the subsequent development of neutrino-driven convection and the SASI, contrasted in the F15.79 model where the low-frequency feature is far more significant and lasts longer, persisting after the onset of explosion. 

\begin{figure}
    \centering
    \includegraphics[width=\columnwidth]{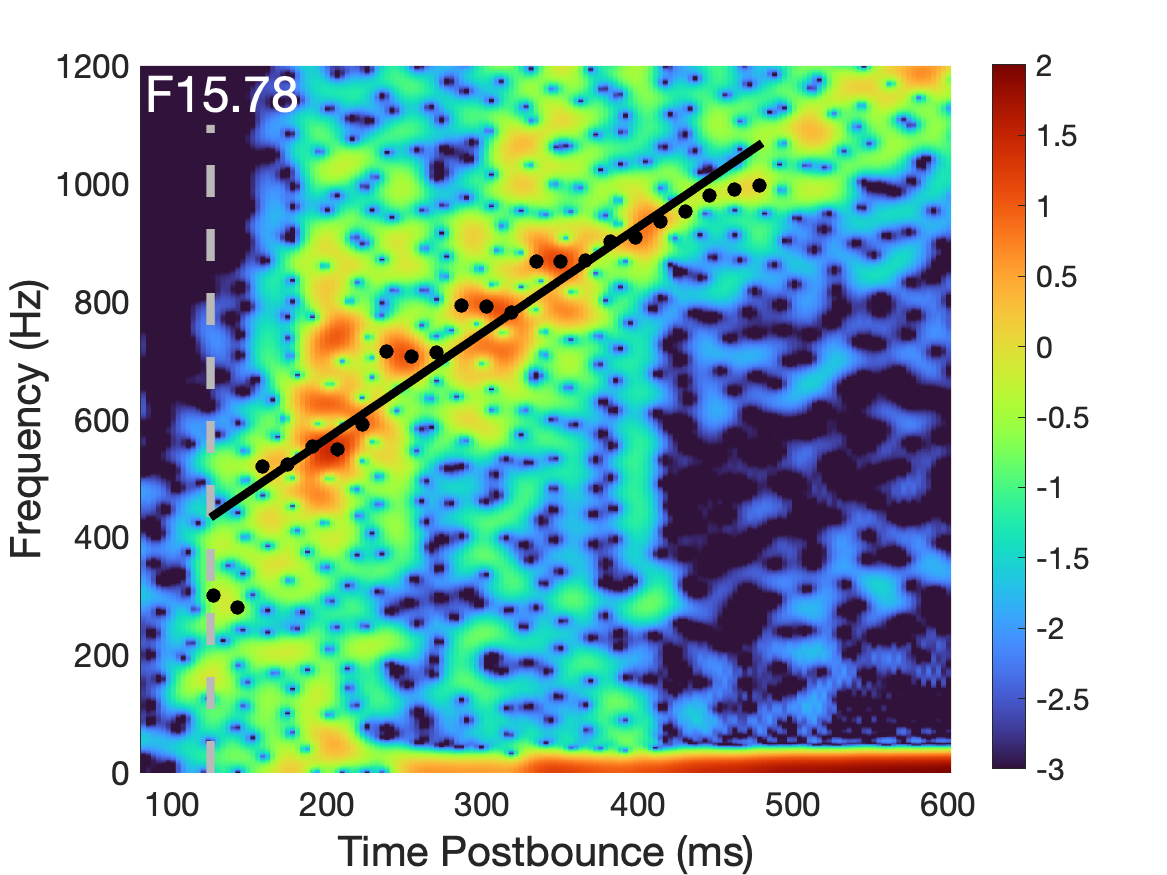}
    \includegraphics[width=\columnwidth]{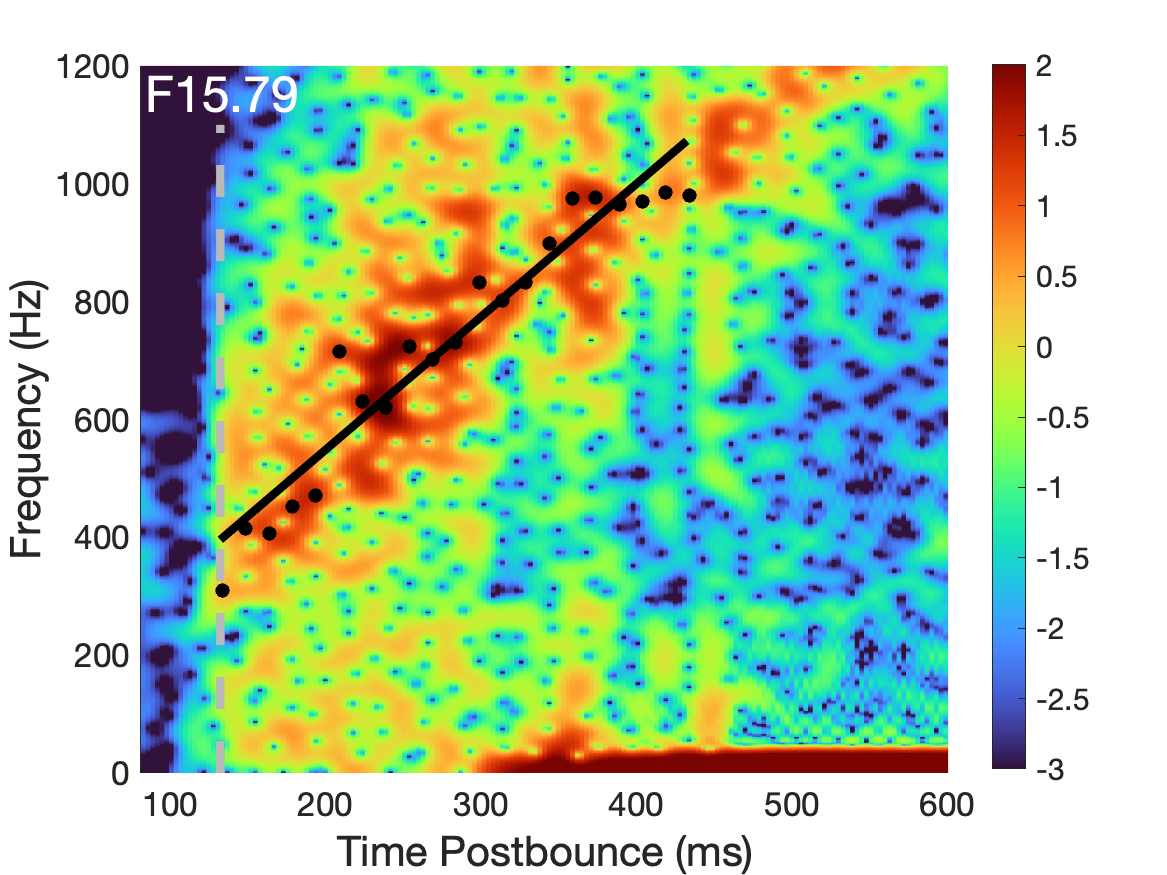}
    \caption{Spectrograms overlaid with the linear approximation of the gfF, for F15.78 and F15.79. The gray dashed line denotes the start time of the gfF.}
    \label{fig:Spec_linear}
\end{figure}

Following the procedure of \citet{MuCaMe24}, we also determine the linear approximation of the peak frequency evolution in the absence of detector noise, shown in Figure \ref{fig:Spec_linear}. In \citet{MuCaMe24}, \citet{CaAnMo23}, and \citet{MuMeLe25pre} the peak frequency evolution is termed the High Frequency Feature (HFF). \citet{MuMeLe25pre} connect this feature to low-order $g$ and $f$ quasinormal modes of oscillation of the PNS. This is in agreement with the analysis of CCSN simulations of other groups \cite{MoRaBu18,RaMoBu19,SoTa20,NaTaKo22,RoRaCh23,ZhAnOc24,SoMuTa24}. For this reason, in this work we will refer to this feature as the $g$-/$f$-mode feature (gfF) to connect it to a physical process rather than an emergent property of the gravitational wave signal.

The linear approximation of the gfF is found by determining the frequencies with maximum power spectral density in each time window, from the beginning of the feature, $t_{\rm gfF}$,  to the time at which the maximal spectral density frequency reaches 1000 Hz. For details on the definition of $t_{\rm gfF}$ (termed $t_{\rm HFF}$ in the reference) and the reason for our cutoff at 1000 Hz, see \citet{MuCaMe24}. The coefficients, starting time for the gfF, and maximum actual percent error (MAPE) of the linear approximation are given in Table \ref{tab:comp}. We find that the slope for F15.78 is 1789 Hz s$^{-1}$ and the slope for F15.79 is 2263 Hz s$^{-1}$. As the PNS quasinormal oscillation modes generate gravitational waves, the radius of the PNS contracts and causes the oscillation frequency to rise. The higher slope of the gfF for F15.79 thus agrees with the greater contraction rate of the PNS seen in Figure 8 of \citet{BrSiLe23}.

\begin{table}
    \begin{center}
        \begin{tabular}{ l c c c c c c } 
            \hline
            \hline
            & $\alpha$ &  $\beta$ & $t_{\rm gfF}$  & MAPE\\
                &  [Hz s$^{-1}$]  &   [Hz]    &   [ms]    & [\%]\\
            \hline
            F15.78 & 1789 & 433.6 & 125.0 & 6.958\\ 
            
            F15.79 & 2263 & 394.9 & 132.5& 6.778 \\
            \hline
        \end{tabular}
        \caption{Linear approximation coefficients for the gfF of the form $f(t)=\alpha t + \beta$. The second column gives the linear slope, $\alpha$, of the gfF. The third column gives the starting frequency of the gfF, $\beta$. The fourth column is the start time of the gfF, $t_{\rm gfF}$. The fifth column represents the maximum absolute percent error (MAPE) between the actual maximum frequencies $A_t$ and the frequencies predicted from the linear approximation $F_t$ for $n$ fitted points: $\rm{MAPE}\equiv100\frac{1}{n}\sum^n_{t=1}|\frac{A_t-F_t}{A_t}|$.}
          \label{tab:comp}
    \end{center}
    \end{table} 

\subsection{Gravitational Wave Luminosity}
\begin{figure}
    \centering
    \includegraphics[width=\columnwidth]{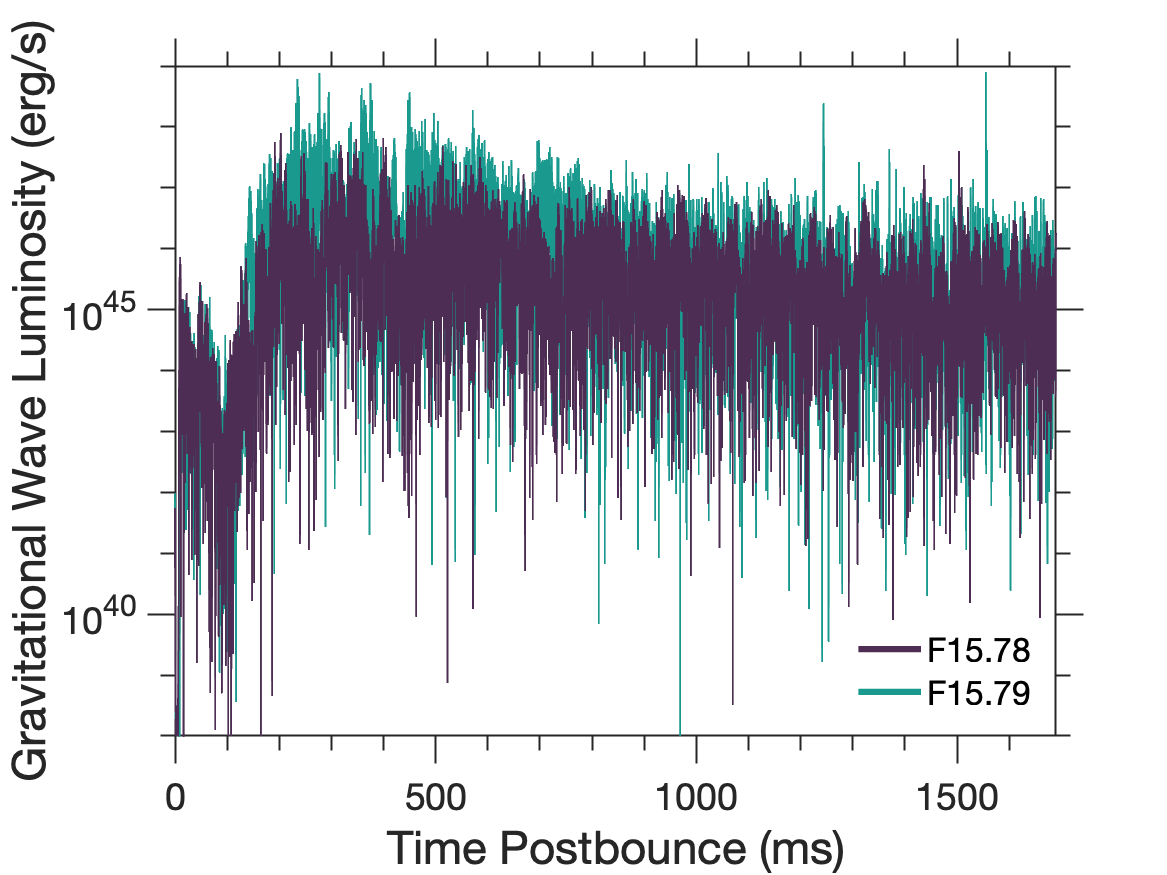}
    \includegraphics[width=\columnwidth]{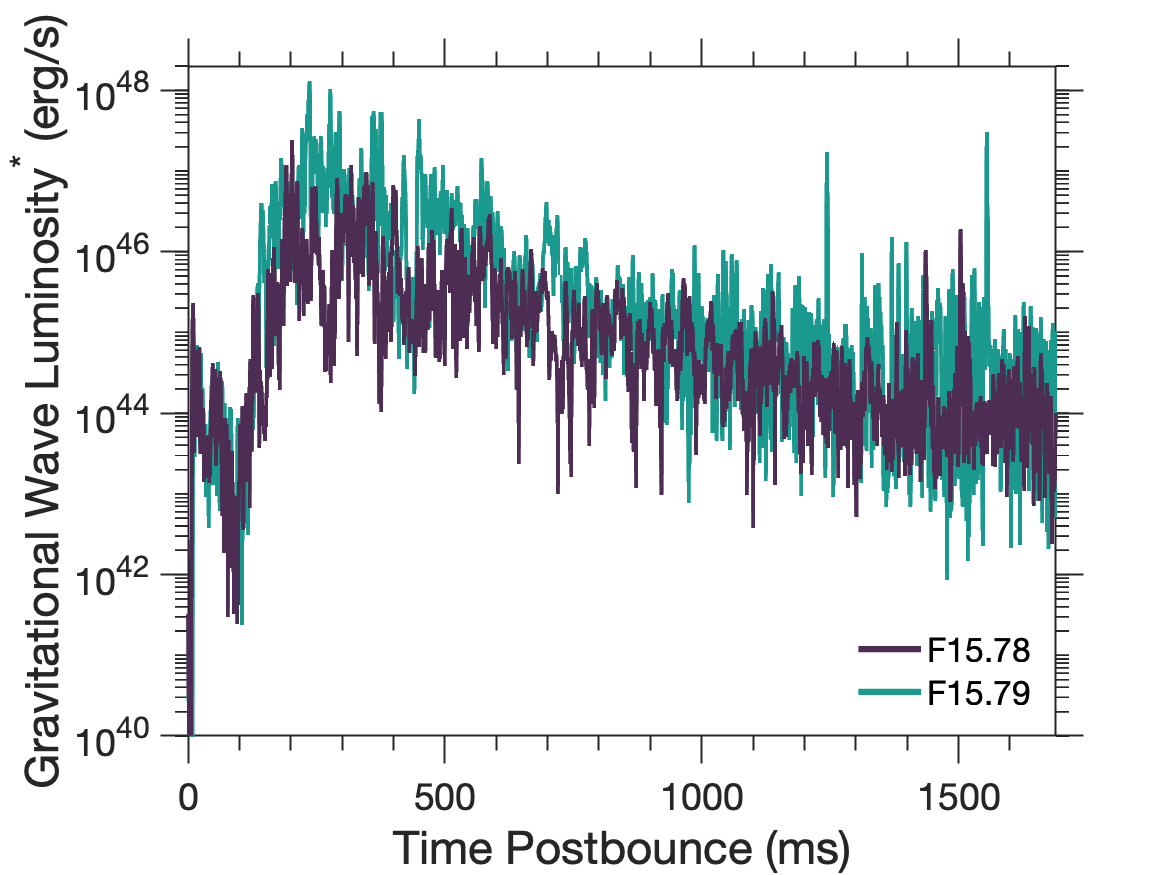}
        \caption{Top: Gravitational wave luminosity for F15.78 (purple) and F15.79 (teal). Bottom: Gravitational wave luminosity computed by averaging over seven time cycles, with the same color labels.}
    \label{fig:lum}
\end{figure}
\begin{figure}
    \centering
    \includegraphics[width=\columnwidth]{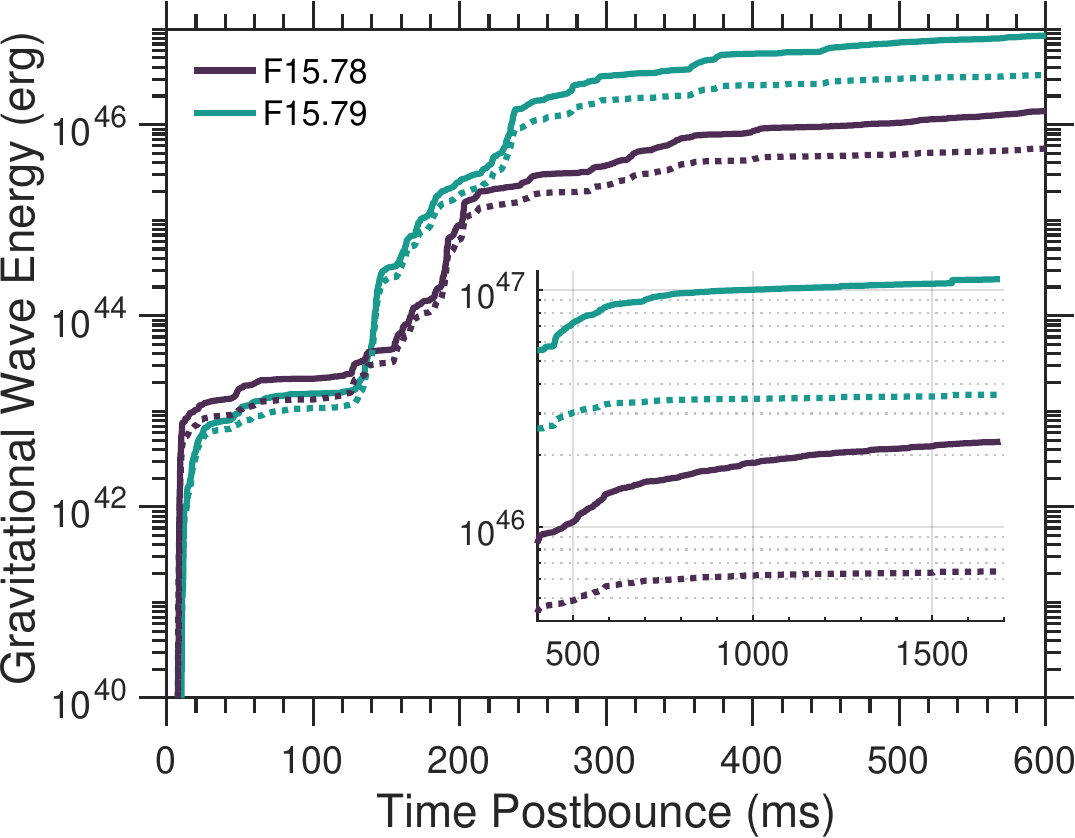}
    \caption{Cumulative gravitational wave energy emitted for F15.78 (purple) and F15.79 (teal), with corresponding energy from the seven-cycle average denoted by dotted lines. The inset shows the gravitational wave energy for late times.}
    \label{fig:energy}
\end{figure}

The gravitational wave luminosity is presented in Figure \ref{fig:lum}, and the cumulative gravitational wave energy is shown in Figure \ref{fig:energy}. As in \citet{MeMaLa23}, the bottom panel of Figure \ref{fig:lum} also includes the gravitational wave luminosity averaged over seven cycles to eliminate high-frequency noise in the emission. This allows for easier distinguishability between the signals, but also lowers the luminosities as well as the emitted energies, the latter of which are indicated with dotted lines in Figure \ref{fig:energy}. We see that the gravitational wave luminosity is greater in F15.78 for the first few 10s of ms after bounce compared to F15.79, but at $\sim$135 ms postbounce F15.79 begins to produce greater gravitational wave luminosities. This is matched by the evolution  of the gravitational wave energy, shown in Figure \ref{fig:energy}, where the total energy emitted as gravitational waves in F15.79 surpasses that of F15.78 at $\sim$140 ms. The bulk of the gravitational wave emission has finished by the end of each simulation.

\section{Summary, Discussion, and Conclusion}
\label{sec:summary}

We computed the gravitational wave strains produced by the CCSNe of two progenitor stars with nearly identical mass but of significantly differing internal structure. The differing explosion dynamics of these two models, as outlined in \citet{BrSiLe23}, resulted in distinct gravitational wave strains, as seen in Figure \ref{fig:strain_total}. We showed that the initial signal due to prompt convection immediately after core bounce is comparable between models, but significant differences are observed beginning $\sim$140 ms after bounce.

We connected the largest gravitational wave strains to accretion events at the surface of the PNS for both models. During the time of maximum accretion onto the PNS, $\sim$150--700 ms as shown by Figure 16 in \citet{BrSiLe23}, we see the maximum gravitational wave strains for both models, and the strains in the more compact F15.79 model are significantly larger than in F15.78. Additionally, the low-frequency component of the gravitational wave strain from explosion and anisotropic neutrino emission is larger in F15.79. 

Our data also agrees with the positive correlation found between compactness parameter and peak frequency evolution found in \citet{WaCoOc20} and \citet{PaWaCo21}. Figure \ref{fig:Spec_linear} shows that the more compact model, F15.79, has a peak frequency that increases more rapidly, as determined by the larger slope of the gfF, than for the less compact model. Table \ref{tab:comp} shows that the difference in slope of the gfF between models is $\sim$26\%, with a difference in starting frequency of $\sim$10\%. As in \citet{MuCaMe24}, we quantify this variation in the absence of any other progenitor differences (e.g., differences in rotation rate). \citet{WaPa24} show the effects of varying progenitor rotation rate in tandem with compactness parameter. They present models where progenitors with relatively higher compactness parameters have a smaller gfF slope than a less compact counterpart that is rotating less rapidly. It is thus clear that precisely determining the interior structure of the progenitor using gravitational waves will be dependent on accurate determination of other progenitor properties.

In our two-dimensional simulations, the differing accretion histories onto the PNS between models account for the significantly differing amplitudes of the gravitational wave strains. It is now well established that the gravitational wave emission in two-dimensional and three-dimensional CCSN simulations differ considerably in both amplitude and sourcing (e.g., see \cite{AnMuMu17,OCCo18,MeMaLa20}). In particular, two-dimensional simulations experience an inverted energy cascade, i.e., energy transfer from small to large scales, that favors the creation of large eddies in post-shock neutrino-driven convection. This means that the accretion onto the PNS in two-dimensional simulations involves fewer, more massive accretion funnels than in three-dimensional simulations. Therefore, three-dimensional simulations, counterparts to the two-dimensional simulations presented here, would have to be performed to ascertain precisely how great the differences between the gravitational wave signals would remain. Nonetheless, there are aspects of the gravitational wave signal that seem to be largely insensitive to the dimensionality of the simulation. The contraction rate of the PNS, and thus the slope of the gfF, is almost identical for the two- and three-dimensional simulations considered by \cite{SoTa20a} and \cite{MuCaMe24}. Thus, this study serves as a proof of principle that gravitational waves cab be used to probe progenitor in internal structure.

\section*{Acknowledgments}

A.M. acknowledges support from the National Science Foundation's Gravitational Physics Theory Program through grants PHY-1806692, PHY-2110177, and PHY-2409148. P.M. is supported by the National Science Foundation through its employee IR/D program. The opinions and conclusions expressed herein are those of the authors and do not represent the National Science Foundation.

This research used resources of the Oak Ridge Leadership Computing Facility at the Oak Ridge National Laboratory, which is supported by the Office of Science of the U.S. Department of Energy under Contract No. DE-AC05-00OR22725. 

\bibliography{apj_journals,pr_add_journals,clean}
\end{document}